\documentclass[11pt,a4paper]{article}
\pdfoutput=1
\usepackage{jcappub}
\usepackage[utf8]{inputenc}
\usepackage[normalem]{ulem}
\usepackage{tikz}
\usepackage{pgf}
\usetikzlibrary{arrows,shapes}
\usepackage{amsfonts}
\usepackage{fullpage}

\def\f{\frac}

\def\ns{n_{\rm s}}
\def\as{A_{\rm s}}

\def\ks{k_*}

\def\ig{\includegraphics}

\def\Q{{q}}
\def\omm{\Omega_{\rm m}}
\def\omr{\Omega_{\rm r}}
\def\oml{\Omega_{\rm \Lambda}}
\def\omk{\Omega_{\rm k}}
\def\lcdm{$\Lambda{\rm CDM}~$}
\def\rhok{\rho_{\rm k}}
\def\rhor{\rho_{\rm r}}
\def\rhophi{\rho_\phi}
\def\phid{\dot\phi}
\def\efolds{{\it e}-folds~}
\def\Pclosed{{\mathcal P}_{\rm closed}}
\def\Pflat{{\mathcal P}_{\rm flat}}
\def\camb{\texttt{CAMB}~}
\def\cosmomc{\texttt{COSMOMC}~}
\def\V0{\mathcal{V}_o}
\def\ro{r_o}
\def\celltt{C^{\rm TT}_\ell}
\def\planck{{\it Planck} }
\def\gone{\gamma^{(1)}}
\def\gtwo{\gamma^{(2)}}

\def\dphi{\delta\phi}
\def\phio{\mathring{\phi}}

\def\dgamma{\delta\gamma}
\def\gammao{\mathring{\gamma}}
\def\Qij{\mathcal S_{ij}^{nlm}}
\def\Pij{\mathcal T_{ij}^{nlm}}
\def\Qupij{\mathcal S^{ij}_{nlm}}
\def\Pupij{\mathcal T^{ij}_{nlm}}
\def\sumn{\sum_{n=2}^{\infty} \sum_{l=0}^{n-1} \sum_{m=-l}^{l}}
\def\mpc{{\rm Mpc^{-1}}}
\def\mpl{m_{\rm Pl}}

\usepackage{enumerate}

\newcommand{\be}{\nopagebreak[3]\begin{equation}}
\newcommand{\ee}{\end{equation}}
\newcommand{\bfig}{\nopagebreak[3]\begin{figure}}
\newcommand{\efig}{\end{figure}}
\newcommand{\ba}{\nopagebreak[3]\begin{eqnarray}}
\newcommand{\ea}{\end{eqnarray}}

\newcommand{\bmult}{\nopagebreak[3]\begin{multline}}
\newcommand{\emult}{\end{multline}}
\newcommand{\Fref}[1]{Fig.\,\ref{#1}}
\newcommand{\fref}[1]{fig.\,\ref{#1}}
\newcommand{\eref}[1]{eq.\,(\ref{#1})}

\newcommand{\sref}[1]{Sec.\,\ref{#1}}
\newcommand{\appref}[1]{appendix\,\ref{#1}}
\title{Inflation in the closed FLRW model and the CMB}

\author[1]{B\'eatrice Bonga,}\emailAdd{bpb165@psu.edu}

\author[2]{Brajesh Gupt,}\emailAdd{bgupt@gravity.psu.edu}

\author[3]{Nelson Yokomizo}\emailAdd{yokomizo@gravity.psu.edu}

\affiliation{
Institute for Gravitation and the Cosmos \& Physics Department, The Pennsylvania State University, University Park, PA 16802 U.S.A.
}
\abstract{
Recent cosmic microwave background (CMB) observations put strong constraints on the spatial curvature via 
estimation of the parameter $\omk$ assuming an almost scale invariant primordial 
power spectrum. We study the evolution of the background geometry and gauge-invariant 
scalar perturbations in an inflationary closed FLRW model and 
calculate the primordial power spectrum. We find that the inflationary dynamics
 is modified due to the presence of spatial curvature, leading to corrections to 
the nearly scale invariant power spectrum at the end of inflation.
When evolved to the surface of last scattering, the resulting temperature
anisotropy spectrum ($\celltt$) shows deficit of power at low multipoles 
($\ell<20$). By comparing our results with the recent \planck data we discuss 
the role of spatial curvature in accounting for CMB anomalies and in the 
estimation of the parameter $\omk$. Since the curvature effects are limited to
low multipoles, the \planck estimation of cosmological parameters
remains robust under inclusion of positive spatial curvature.
}

\preprint{}

\begin{document}
\maketitle

\section{Introduction}

The recent measurements of the cosmic microwave background radiation (CMB) anisotropies by
observational missions such as \planck \cite{Ade:2015lrj} and WMAP \cite{wmap9}
are in excellent agreement with the well known \lcdm model of the Universe.
According to this model, the geometry of spacetime is described by a homogeneous 
and isotropic Friedmann-Lema\^itre-Robertson-Walker (FLRW) metric, and the energy 
content of the  Universe is composed of baryonic and cold dark matter ($\omm$), 
radiation ($\omr$), a cosmological constant ($\oml$) and spatial curvature ($\omk$) 
\cite{Ade:2015lrj,wmap9,weinberg,mukhanov,dodelson}. Within this scenario, the 
observational data imposes strong constraints on the spatial curvature of the 
FLRW background: $\omk=-0.005_{-0.017}^{+0.016}$ from \planck measurements alone, and 
$\omk=0.000_{-0.005}^{+0.005}$ when the \planck measurements are combined 
with BAO data \cite{Ade:2015xua} (results are quoted at 95\% confidence intervals). 
The origin of the tiny temperature fluctuations in the CMB can be attributed to 
almost scale invariant primordial density perturbations predicted by inflation. 
In the simplest inflationary scenario the accelerated expansion of the Universe 
is driven by a single scalar field $\phi$ with standard kinetic energy 
slowly rolling down a potential $V(\phi)$ in a spatially flat ($\omk=0$) 
FLRW spacetime \cite{guth,albrecht,linde,lindechaotic,liddlelyth,mukhanov,weinberg}.

While consistent with a spatially flat spacetime geometry, the current bounds on 
$\omk$ allow for the possibility that we live in a slightly curved patch of the 
Universe, which can be related to a global non-zero spatial curvature, e.g. open 
or closed FLRW spacetime, or to the effect of super-horizon fluctuations on our 
local patch in a globally flat Universe \cite{Ade:2015xua}. Closed FLRW model
has also been argued to be of phenomenological \cite{White:1995qm} and
fundamental importance \cite{Gratton:2001gw}. In the presence of 
spatial curvature, the predictions of inflationary models are affected in two 
distinct ways. First, the transfer functions encoding the evolution of linear 
perturbations from the end of inflation until today are modified by the presence 
of curvature terms in the Boltzmann equations. This effect is well understood 
and is fully implemented in the Boltzmann codes employed for the integration of 
these equations (as \camb \footnote{http://camb.info}, for instance 
\cite{Lewis:1999bs}). Second, the spectrum of primordial perturbations can be 
affected by the presence of curvature during inflation. Recall that the rapid 
accelerated expansion of the Universe during inflation `washes away' any spatial 
curvature that might have been present before inflation, making the spatial 
geometry at the end of inflation very close to flat. Therefore, even if the 
spatial curvature $\omk$ is very small at the end of inflation and today, 
it might have been significant during the early stages of inflation. 
In this scenario, the evolution of quantum
perturbations in a spatially curved spacetime is distinct from that in 
a spatially flat spacetime during inflation, especially near the onset of
inflation. As a result, the spectrum of primordial fluctuations will be 
influenced by the effects of $\omk\neq0$. Significant deviations from the simple 
power law form generally can occur for wavelengths comparable to the curvature 
radius, which are later imprinted as scale-dependent features in the CMB power 
spectra. This effect, however, has not been analyzed in great detail, even for 
small spatial curvatures, $\omk$. In particular, these potential features in the
primordial power spectrum are not taken into account for the estimation of 
cosmological parameters including $\omk$ in \planck analysis 
\cite{Ade:2015xua,Ade:2013zuv}, which is based on a tilted power law spectrum 
for the primordial scalar perturbations. The goal of this paper is to revisit 
these issues for the closed FLRW model, i.e. with $\omk<0$, by studying the
evolution of quantum perturbations in a closed inflationary Universe.

In a closed FLRW model the topology of the spacetime is $\mathbb R\times
\Sigma$ with the topology $\Sigma$ of the spatial sections that of a
3-sphere $\Sigma=\mathbb S^3$. In this case, the angular scales in the CMB 
that are potentially affected by the global curvature can be estimated by 
setting $\omk$ to the observed values. Using the \planck estimate for the Hubble 
constant $H_0=(67.8\pm0.9)$ km s$^{-1}$ Mpc$^{-1}$ and $\omk=-0.005$ 
\cite{Ade:2015xua}, we find that the radius of the spatial $\mathbb S^3$ section 
of the Universe today is: $r_0\simeq 4.5~r_{LS}$, where $r_{LS}\approx 14 
~{\rm Gpc}$ is the radius of the last scattering surface.\footnote{Note that 
$r_0$ is the radius of a 3-sphere representing the whole Universe and $r_{LS}$ 
is radius of the 2-sphere corresponding to the surface of last scattering.} 
This indicates that curvature effects, if present, will be more pronounced at 
length scales comparable to that of the CMB sky. These scales correspond to the 
lowest multipoles. Interestingly, recent CMB observations reveal anomalies in the
temperature anisotropy spectrum at low multipoles including a power
suppression for $\ell\lesssim 30$ that is mildly incompatible with
the predictions of an almost scale-invariant primordial spectrum
\cite{Ade:2015lrj,Schwarz:2015cma,Hunt:2015iua}. Therefore, it is natural to ask 
to what extent spatial curvature can affect the low-$\ell$ spectrum and account 
for this anomaly. 

The power suppression at low multipoles has been studied in the case of open 
inflation, with the help of toy models, by introducing an early period 
of fast-roll at the onset of inflation following the process of bubble 
nucleation \cite{Linde:2003hc,White:2014aua}. A similar mechanism involving a 
transient regime of fast-rolling has also been advocated in flat models 
\cite{Contaldi:2003zv,Jain:2008dw,Pedro:2013pba,Lello:2013awa,Lello:2013mfa,
Cook:2015hma}, showing that the presence of curvature may not be essential in 
this picture. In both cases, however, one needs to include fine-tuned localized 
features in the potential energy $V(\phi)$ of the scalar field on scales 
corresponding to the present day Hubble horizon. In this paper, we focus on the 
less studied case of a closed Universe for the quadratic and Starobinsky potentials 
without extra features.

Due to its finite size, a closed Universe has a characteristic length scale 
given by the curvature radius $r_0$. It is generally expected that the 
primordial fluctuations truncate on such a scale, which thus acts as a natural 
infrared cutoff \cite{Uzan:2003nk,Ellis:2001ym,Ellis:2001yn,Efstathiou:2003hk,
Lasenby:2003ur,Luminet:2003dx}. Under this assumption, it was shown in 
\cite{Efstathiou:2003hk} that the quadrupole anomaly can be consistently interpreted 
as a curvature effect. Evidence for such a behavior was presented in 
\cite{Masso:2006gv}, where it was found that the power spectrum of scalar 
perturbations on a closed de Sitter background is slightly suppressed at the 
scale of the curvature radius. This raises the question 
as to whether the full inflationary dynamics can actually enforce the conjectured
truncation and how exactly the primordial power spectrum departs from the 
power law characteristic of flat spaces at the length scale set by $r_0$. 

In order to investigate these questions, we perform a detailed analysis of the 
inflationary evolution of gauge-invariant scalar perturbations in a closed FLRW 
Universe for a scalar field with a quadratic and Starobinsky potential, 
based on the Hamiltonian formalism of \cite{Langlois1994}. We first determine 
initial conditions for the background geometry which lead to a sufficiently long 
slow-roll phase that is compatible with observations. Note that it is a priori not 
clear if a given initial condition will lead to inflation at all 
\cite{Barrow:1988xi}. Providing initial conditions for the 
perturbations at the onset of inflation, we numerically integrate these equations 
of motion for a range of values of $\omk$ including the \planck estimates, 
$\omk =-0.005$. Since the curvature term in the Friedmann equation scales as 
$a^{-2}$, where $a$ is the scalar factor, the background evolution is strongly 
affected by the spatial curvature in the early stages of inflation, leading to a 
breakdown of the slow-roll approximation \cite{Uzan:2003nk}. This mechanism 
allows for a transient fast-roll regime without the need of modifying the 
potential, with significant implications for the modes crossing the horizon at 
these times. We calculate the power spectrum at the end of inflation and
use the Boltzmann code \camb to determine the spectrum of temperature anisotropies 
in the CMB. We find that the modifications in the primordial power spectrum
lead to power suppression at multipoles $\ell\lesssim20$, with almost 30\%
suppression at the lowest multipole, $\ell=2$. This indicates that, while
this suppression of power is not strong enough to completely explain the 
observed anomaly, the presence of spatial curvature can partially account for 
it. Also, we find that the modification in the power spectrum has
negligible effects on the estimation of cosmological parameters. This happens
because the curvature effects on the CMB are limited to only low $\ell$'s while for
$\ell>20$ the spectrum remains practically unaffected. Therefore, while 
a priori being conceptually inconsistent, the \planck estimation of cosmological 
parameters in the presence of spatial curvature \cite{Ade:2015xua} is 
phenomenologically robust.

The paper is organized as follows. In section \ref{sec:prel} we provide a brief
overview of the equations of motion of the closed FLRW model in the presence of a
scalar field with a potential $V(\phi)$ and gauge-invariant scalar perturbations
on it. Section \ref{sec:inidata} provides details on the initial conditions
for the background variables and quantum perturbations chosen such that the future 
evolution has a minimal duration of inflation and leads to a power spectrum 
compatible with observations at large $\ell$'s. We study the evolution of the 
background spacetime and scalar perturbations in section \ref{sec:results}. This is 
where we compute the temperature anisotropy spectrum and discuss the effects of 
spatial curvature on the CMB spectrum and estimation of cosmological parameters.
We conclude in section \ref{sec:disc} with a summary of the main results and future 
outlook. Technical details about the properties of the spherical harmonics are given in the appendix \ref{app:A}.

%%%%%%%%%%%%%%%%%%%%%%%%%%%%%%%%%%%%%%%%%%%%%%%%%%%%%%%%%%%%%%%%%%
\section{Preliminaries}
\label{sec:prel}
This section is divided into two subsections. In the first subsection we provide 
a brief overview of the background spacetime dynamics of a closed FLRW spacetime
in the presence of a non-minimally coupled scalar field with a self-interacting potential. 
In the second subsection, we discuss the construction of gauge-invariant linear scalar
perturbations $\Q$ using Hamiltonian methods described in \cite{Langlois1994}, 
obtain the second order Hamiltonian
\cite{Halliwell:1984eu,Langlois1994,FernandezMendez:2012sr} and derive the
equations of motion for $\Q$. 

\subsection{Background dynamics}
For the background, we consider a homogeneous, isotropic spatially curved 
Friedmann-Lema\^{i}tre-Robertson-Walker model with topology $\mathbb{R} \times \mathbb{S}^3$. The spacetime metric is 
\begin{equation}
ds^2 = - dt^2 + \tilde{a}^2 (t) \, r_o^2 \, d\Omega^2,
\end{equation}
where $r_o$ is the curvature radius of the $\mathbb S^3$ spatial section, $\Omega_{ij}$ 
is the metric on the fiducial unit three-sphere with volume $\V0= 2\pi^2$ and 
$\tilde a$ is dimensionless scale factor. In spherical coordinates 
$(\chi, \theta, \varphi)$, the fiducial three-sphere metric representing the
spatial section takes the form:
\begin{equation}
d\Omega^2 = d\chi^2 + \sin^2\chi \, \left(d\theta^2 + \sin^2 \theta \, d \varphi^2 \right).
\end{equation}
It is convenient to define the dimensionful scale factor: $a(t):= \ro \,
\tilde{a}(t)$. Given a matter field with the energy density $\rho$ and pressure
$P$, the dynamics of $a(t)$ is given by the Friedmann and Raychaudhuri
equations: 
\begin{eqnarray}
H^2 &=& \frac{8 \pi G}{3} \rho - \f{1}{a^2},
\label{eq:friedmann} 
\\
\dot{H} &=& - 4 \pi G \left( \rho + P \right) + \f{1}{a^2},
\label{eq:raychaudhuri}
\end{eqnarray}
where $H=\dot{a}/a$ is the Hubble parameter. We are interested in the
inflationary dynamics which is driven by a scalar field with a self-interacting 
potential $V(\phi)$. The energy density $\rho$ and the pressure $P$ of the 
scalar field are:
\be
 \rho= \f{1}{2} \dot{\phi}^2 + V(\phi) \qquad {\rm and } \qquad P= \f{1}{2}
\dot{\phi}^2 - V(\phi). 
\ee
Using the above definitions and equations \eqref{eq:friedmann} and
\eqref{eq:raychaudhuri} we obtain the Klein-Gordon equation which governs the
evolution of the scalar field:
\begin{equation}
\ddot{\phi} + 3 H \dot{\phi} + \frac{dV}{d\phi} = 0 \, .
\label{eq:klein-gordon}
\end{equation}
Among the eqs.~\eqref{eq:friedmann}, \eqref{eq:raychaudhuri} and 
\eqref{eq:klein-gordon} only two are independent: any of these equations can be
derived by combining the other two. Therefore, it is sufficient to consider the 
Klein-Gordon equation (\ref{eq:klein-gordon}) together with the  Friedmann
equation (\ref{eq:friedmann}) to completely describe the evolution of the 
background spacetime. These two equations form a well-posed initial value problem,
which --- given proper initial conditions at an initial time --- can be solved to 
evaluate the quantities $a(t)$ and $\phi(t)$ at a later (or earlier) time. In 
this paper, we consider two different potentials for the inflationary phase:
the quadratic potential
\be
V(\phi)=\f{1}{2} m^2 \phi^2 \, ,
\label{eq:quadratic}
\ee
and the Starobinsky potential \cite{Barrow:1988xh,Maeda:1988ab,Starobinsky:2001xq,DeFelice:2010aj}
\be
V(\phi) = \f{3M^2}{32 \pi G} \left(1- e^{-\sqrt{\f{16\pi G}{3}} \phi} \right)^2. \, 
\label{eq:starobinsky}
\ee
In both cases the values of the mass parameters $m$ and $M$ are fixed by using 
Einstein's equations in conjugation with recent observations as described in 
section \ref{sec:inidata}. 

After inflation ends, the scalar field decays completely and standard model 
particles are created during the reheating phase. The further evolution of the 
background spacetime is described by the standard \lcdm model: the matter 
content of the Universe consists of baryonic and cold dark matter 
($w:=P/\rho=0$), radiation ($w=1/3$) and the cosmological constant or dark energy 
($w=-1$). The Friedmann equation describing the background geometry in 
this phase all the way till today is:
\be
  H^2 = H_0^2 \left(\omm~\left(\f{a}{a_0}\right)^{-3}+\omr~\left(\f{a}{a_0}\right)^{-4}+\oml + \omk~\left(\f{a}{a_0}\right)^{-2}\right),
\label{eq:friedlcdm}
\ee
where $H_0$ is the Hubble parameter today. The parameters $\omm=0.31,~\omr=9.2\times10^{-5},~\oml=0.69$ and $\omk$ are the contribution to the total extrinsic 
curvature from matter, radiation, dark energy and spatial curvature, 
respectively, whose values are measured today by the recent CMB 
experiments. For the closed FLRW model under consideration, $\omk<0$. The 
\planck power spectra constrain 
$\omk=-0.005^{+0.016}_{-0.017}$ at a 95\% confidence interval.  Adding data 
from BAO to break the `geometric degeneracy' between $H_0$ and $|\omk|$, yields 
$\omk=0.000^{+0.005}_{-0.005}$ at the same confidence interval \cite{Ade:2015xua}.

\subsection{Perturbations}
On the homogeneous and isotropic inflationary background described above, we 
introduce first order, purely inhomogeneous perturbations:\footnote{Just as in 
the flat Universes, the scalar, vector and tensor modes decouple at first order
and evolve independently of one another. In this paper, we only consider 
scalar perturbations.}  
\be
   \gamma_{ij} = \gammao_{ij} + \dgamma_{ij}; \qquad      
    \phi = \phio + \dphi,
\ee
where $\gammao_{ij}=a^2\Omega_{\ij}$ is the background metric, $\phio$ is the background homogeneous inflaton field, and $\dgamma_{ij}$ and $\dphi$ are the inhomogeneous metric
and inflaton perturbations respectively. The metric perturbation $\dgamma_{ij}$
has two degrees of freedom, let's denote them $\gone$ and $\gtwo$ (see \appref{app:A} for exact relation), which are equivalent to the two Bardeen potentials (usually denoted by $\Psi$ and $\Phi$ in the literature \cite{Bardeen:1980kt,mfb_review}). 
Paying attention to the symmetries of the spatial hypersurface we expand these 
perturbations in scalar harmonic functions $Q_{nlm}$ and tensor harmonic functions $\Pij$ and $\Qij$ (defined in \appref{app:A}), which form a complete basis on $\mathbb{S}^3$ \cite{Halliwell:1984eu,Langlois1994,FernandezMendez:2012sr}:
\ba
\dgamma_{ij} &=& a^2 6\sqrt{\mathcal V_o} \sumn 
                  \left(\gone_{nlm} \Qij + \gtwo_{nlm} \Pij\right)  
\label{eq:metric-perturb}
   \\
\delta \phi &=& 
\sum_{n=2}^{\infty} \sum_{l=0}^{n-1} \sum_{m=-l}^{l} \delta \phi^{nlm} Q_{nlm}.
\label{eq:field-perturb}
\ea
Note that the sum starts at $n=2$ since the $n=1$ mode is homogeneous and thus part 
of the background. These harmonics are scalar eigenfunctions of the Laplacian operator 
on $\mathbb{S}^3$: $\nabla^2 Q_{nlm} = -(n^2 -1) Q_{nlm}$ with $\nabla$ the covariant 
derivative compatible with $\Omega_{ij}$. Their explicit form is  
\begin{equation}
\label{eq:harmonics}
Q_{nlm} (\chi,\theta, \varphi) = \Phi_n^l (\chi) \, Y_{lm}(\theta, \varphi) \, ,
\end{equation}
where $Y_{lm}$ are the spherical harmonics and the expression for the `radial' 
function $\Phi_n^l$ can be found in appendix \ref{app:A}. This is analogous to 
the flat model, in which any scalar function can also be expanded in terms of the 
eigenfunctions of the flat space Laplacian with eigenvalue $-k^2$. We will 
denote the flat space eigenfunctions by $Q_{klm}$. The structure of the eigenfunctions 
$Q_{klm}$ is similar to \eref{eq:harmonics}, where the radial function only depends 
on $l$ and is simply the spherical Bessel function $j_l(kr)$. The curved 
$\Phi_n^l(\chi)$ tends to $j_l(kr)$ (with $n\chi \to kr$)
in the limit of large $n$ and $\chi \ll 1$, which is exactly the limit from 
the closed to flat FLRW model. In this limit, the hyperspherical harmonics 
of the closed model $Q_{nlm}$ reduce to the spherical harmonics of the
flat model $Q_{klm}$. By comparing the eigenvalues of the \textit{physical} 
Laplacian in both cases today, we find that:
\begin{equation}
\frac{n^2-1}{r_o^2} \rightarrow {k^2} \, ,
\label{eq:n-k}
\end{equation}
where $r_o$ is the radius of the closed Universe today and $k$ is the comoving
wavenumber in the flat limit. The factor $n^2-1$ is typically interpreted as the 
curved version of $k^2$. In fact, it is convenient to introduce $k$ even in a 
closed Universe --- where plane waves cannot be defined --- using the expression above, 
in order to have an instantaneous notion of wavelength in terms of the labels $n$ 
\cite{Lewis:1999bs}. The radius $r_o$ depends on the parameters $\omk$
and $H_0$ as: 
\be
r_o = \f{1}{\sqrt{|\omk|} H_0}.
\ee 
Using $\omk=-0.005$ and {\it Planck's} estimate for the Hubble parameter 
$H_0=67.3~{\rm km~s^{-1}~Mpc^{-1}}$ we find that 
$r_o\approx 6.3\times10^4~{\rm Mpc}$. For such a Universe, the \planck comoving 
pivot scale $k=0.05~\mpc$ corresponds to  $n=3152$, and $k=0.002~\mpc$  
corresponds to $n=126$.

As in the flat model, the first order scalar perturbations 
($\gone,\gtwo,\dphi$) are gauge dependent quantities, and only certain
combinations of them are gauge-invariant. The gauge-invariant variable $\Q$ 
was obtained using methods developed for first-class Hamiltonian constrained 
systems \cite{Goldberg1991}, which were applied to the cosmological setting by 
Langlois \cite{Langlois1994}. These methods allow for the determination of a 
singular transformation from the constraint surface in phase space to the 
reduced phase space of gauge-invariant, physical degrees of freedom. 
The procedure is as follows. First, general relativity is cast in the Hamiltonian 
formalism. The phase space relevant for cosmological perturbation theory is 
$\Gamma_o \times \Gamma_1$, with $\Gamma_o$ the 4-dimensional phase space of the 
unperturbed background, homogeneous fields and $\Gamma_1$  phase space of 
the first order, purely inhomogeneous perturbations $(\gone,\gtwo,\dphi)$ and 
their conjugate momenta with six scalar degrees of freedom. 
Next, the first order Hamiltonian and momentum constraints select a surface 
$\Gamma_1^{c} \subset \Gamma_1$ with four (scalar) degrees of freedom. These 
constraints, being first class, also generate gauge transformations within 
$\Gamma_1^{c}$, which further reduce the number of degrees of freedom to two, 
i.e. a single gauge-invariant canonical pair $(q,p) \in \tilde{\Gamma}_1 \subset \Gamma_1^c$. To 
obtain the gauge-invariant perturbation, one needs to use the first order 
Hamiltonian and momentum constraints to go from $\Gamma_1$ to the reduced 
phase space $\tilde{\Gamma}_1$, defined as the set of gauge orbits in the 
constraint surface $\Gamma_1^c$. This is done using 
Hamiltonian-Jacobi inspired equations, which select a generating function 
that nicely implements a singular coordinate transformation from  
$\Gamma^c_1$ to $\tilde{\Gamma}_1$. As a result one obtains the following
expression for the gauge-invariant variable $\Q$ \cite{Langlois1994}:
\be
 \Q_{nlm} = \dphi_{nlm} -\sqrt{\mathcal V_0} \f{\phid}{H}\left(\gone_{nlm} + \gtwo_{nlm}\right),
\ee
where $q$ is written in terms of hyperspherical harmonics as $\Q = \sum_{n=2}^{\infty} \sum_{l=0}^{n-1} \sum_{m=-l}^{l} \Q_{nlm} Q_{nlm}$.
The gauge-invariant variable $q$ is proportional to the usual curvature perturbation 
$\zeta$: $\Q = (\dot{\phi}/H) \zeta$.\footnote{In terms of the scalar field perturbation 
$\delta \phi$ and Bardeen potential $\Psi$, this gauge-invariant quantity is 
$q = \delta \phi + \frac{\dot{\phi}}{H} \Psi$ \cite{mfb_review}.} The dynamics
of $\Q$ on the reduced phase space is then generated by the second order
Hamiltonian constraint in which only terms quadratic in the first order
perturbations are kept. This yields a total quadratic Hamiltonian which is then
used to obtain the following equation of motion for $\Q_{nlm}$:
\begin{equation}
\label{eq:evolutionQ}
\ddot{\Q}_{nlm} + b(n, t) \dot{\Q}_{nlm} + c(n, t) \Q_{nlm} = 0 
\end{equation}
where for each mode $b$ and $c$ are completely determined by the background evolution:
\begin{multline}
b(n, t)  = 3 H + \f{32 \pi  G a^3 \dot{a} \dot{\phi} V'(\phi) 
              +48 \pi  G a^2 \dot{a}^2 \dot{\phi}^2   
         -8 \pi  G a^2 \dot{\phi}^2 \left(8 \pi  G a^2 \left(\dot{\phi}^2-2 V \right)
               +2\right)}{2 a \dot{a} \left(2(n^2-4)\dot{a}^2 
              + 8 \pi G a^2 \dot{\phi}^2 \right)} 
%              \Bigg[ 32 \pi  G a^3 \dot{a} \dot{\phi} V'(\phi) 
%              +48 \pi  G a^2 \dot{a}^2 \dot{\phi}^2   \\
%         -8 \pi  G a^2 \dot{\phi}^2 \left(8 \pi  G a^2 \left(\dot{\phi}^2-2 V \right)
%               +2\right) \Bigg]
\end{multline}
\begin{multline}
 c(n, t)  =  \f{8\pi G}{ a^2 \dot{a}^2 \left(2 \left(n^2-4\right) 
                \dot{a}^2+8 \pi G  a^2 \dot{\phi}^2\right)} \\ 
         \Bigg[ \f{\dot{a}^4 (n^2-4) \left( n^2-1 + a^2 V'' \right)}{4\pi G}+ \left(4 n^2-7\right) a^3 \dot{a}^3 \dot{\phi} V' -\pi G \frac{n^2-1}{n^2-4} a^4\dot{\phi}^4 
         \left[8 \pi  G a^2 \left(\dot{\phi}^2+2 V\right)-6\right] \\
      +   (n^2-1) a^2 \dot{a}^2  \Bigg(  
            -6 \pi  G  \frac{n^2-5}{n^2-4} a^2\dot{\phi}^4+ 
             4 \pi G a^2 \dot{\phi}^2 V+\frac{3}{2} \dot{\phi}^2 + \frac{9}{2} \dot{a}^2 \dot{\phi}^2 \Bigg) \\
                      + a^3 \dot{a} \left[ a \dot{a} \dot{\phi}^2 V''+2 a \dot{a} V'^2+ 4 \pi G a^2 \dot{\phi} V' \left(\dot{\phi}^2+2 V\right)- \dot{\phi} V'  \right] \Bigg]
\end{multline}
with a prime denoting a derivatives with respect to $\phi$.
A few comments are in order concerning the evolution of $\Q^{nlm}$: 
\begin{itemize}
\item Given an inflationary potential and proper initial conditions chosen at some
      initial time, \eref{eq:evolutionQ} describes the evolution of $\Q_{nlm}$
      all the way till the end of inflation. 
\item It is immediately obvious that the equation of motion decouples for each mode, which 
      allows one to solve the evolution of the perturbations mode by mode. This is not 
      surprising as the same is true in the absence of spatial curvature. 
\item In the limit of large radius $r_o$ and large $n$, \eqref{eq:evolutionQ} reduces 
      to the familiar evolution equation in the case of flat spatial curvature 
      \cite{mfb_review}. Moreover, the divergent behavior for $n=2$ in 
      \eqref{eq:evolutionQ} is not a problem, because the $n=2$ mode for $\Q$ 
      vanishes identically. 
\item If the background is given by a de Sitter geometry, then $b(n, t) = 3H$ and 
      $c(n, t) = - a^{-2} (n^2-1)$.
\end{itemize}
Having the equations of motion for both the background and gauge-invariant
linear perturbations, we are set to study their evolution during inflation. Let
us now discuss the choice of initial conditions for the background and
perturbations.\footnote{In the post inflationary phase, during the radiation and matter domination era,
the evolution of cosmological perturbations is governed by the Boltzmann
equations which is implemented in the publicly available code
\texttt{CAMB} \cite{Lewis:1999bs}.}

\section{Initial conditions}
\label{sec:inidata}

The dynamics of the background during the inflationary era is described by 
Eqs.~\eqref{eq:friedmann}--\eqref{eq:klein-gordon}. The initial data required for solving this system of evolution equations consists of a choice of initial conditions 
$(a_0,\phi_0,\dot{\phi}_0)$. In addition, one must fix the value of the mass 
parameter in the potential $V(\phi)$ of the inflaton. In our analysis, we have 
considered a space of initial data and inflaton masses restricted by the condition 
that background solutions must include an inflationary regime
%, characterized by an accelerated expansion, $\ddot{a}>0$, 
long enough so that all observable modes in 
the CMB are within the comoving Hubble horizon at the onset of the inflation. The 
onset of inflation $t_0$ is defined as the instant at which $\ddot{a}(t_0)=0$ and 
the comoving Hubble horizon assumes its maximal value. Explicitly, we adopt the 
condition:
\be
\lambda_{\rm max} \lesssim \frac{r_o}{a(t_0) H(t_0)} \, ,
\label{eq:long-enough-inf}
\ee
where $\lambda_{\rm max}$ is the comoving wavelength of the largest observable mode. 
Let us describe the procedure adopted for implementing this condition.

Consider first the case of a quadratic potential \eqref{eq:quadratic}, for definiteness. In order to fix the mass parameter, we first assume that: (i) the Universe was approximately flat at the time $t_\ast$ when $k_\ast=0.05$ Mpc$^{-1}$ exited the horizon, and (ii) the slow-roll approximation is valid at $t_\ast$. Under these approximations, we can use the usual expressions for the amplitude of the scalar power spectrum $\as$ and its running $\ns$ in terms of background quantities to obtain:
\be
\epsilon = \frac{1-\ns}{4}  \, , \qquad H_\ast = \sqrt{\frac{\pi \as
\epsilon}{G}} \, ,
\label{eq:ns-as}
\ee
where $\epsilon =-\dot H/H^2$ is the first slow-roll parameter. The scale factor at 
horizon crossing is by definition equal to $a_\ast/r_o = k_\ast/H_\ast$, leading to a formula for $a_\ast$ in terms of the observable quantities $\as$ and $\ns$:
\be
a_\ast = 2 r_o k_\ast \sqrt{\frac{G}{\pi \as (1-\ns)}} \, .
\label{eq:a-ast}
\ee
Moreover, the slow-roll parameter is related to the field configuration by
\be
\dot{\phi_\ast}^2 (\epsilon-3) + \epsilon ~ m^2 \phi_\ast^2 = 0 \, ,
\label{eq:epsilon-phi}
\ee
and, under the assumed approximations, the background dynamics is described by:
\begin{gather}
3 H \dot{\phi} + m^2 \phi = 0 \, , 
\label{eq:kg-slow-roll}
\\
H^2 = \frac{8 \pi G}{3} \left( \frac{\dot{\phi}^2}{2} + \frac{m^2 \phi^2}{2}  \right).
\label{eq:friedmann-flat}
\end{gather}
Combining Eqs.~\eqref{eq:epsilon-phi}, \eqref{eq:kg-slow-roll} and \eqref{eq:friedmann-flat}, we can solve for the mass and the field configuration at $t_\ast$:
\be
\phi_\ast =\frac{-3+(1-\ns)/4}{3\sqrt{\pi G (1-\ns)}} \, , \qquad
\dot{\phi}_\ast =\frac{\sqrt{\as} (1-\ns)}{8 G} \, , \qquad
m = \frac{3}{2} \frac{(1-\ns) \sqrt{\pi \as}}{\sqrt{G(11+\ns)}} \, , 
\label{eq:bg-initial-cond-m}
\ee
Allowing the parameters $\as$ and $\ns$ to vary within the $2\sigma$ region determined by the Planck analysis, we obtain a window $\mathcal{I}=\{(a_\ast,\phi_\ast,\dot{\phi}_\ast),m\}$ of initial data and inflaton masses $m$ compatible with the observed data. Each point in this space of allowed initial conditions determines a specific evolution of the background, for which the condition \eqref{eq:long-enough-inf} can be explicitly checked. Our analysis is based on background solutions in the observationally selected window $\mathcal{I}$: 
\ba
\phi_\ast=2.99 \pm 0.52~\mpl, \qquad \phid_\ast = (-2.08 \pm
0.73)\times10^{-7}~\mpl^2, \\ 
a_\ast=(3.3\pm0.6)\times10^{-54}~r_o, \qquad m= (1.28 \pm 0.45) \times 10^{-6}~\mpl,
\label{eq:quadini}
\ea
 such that the condition \eqref{eq:long-enough-inf} is satisfied, where $\mpl=\sqrt{\hbar c/G}$ is the Planck mass.

The assumptions (i) and (ii) can be explicitly verified for solutions in $\mathcal{I}$. In order to do so, we first compare the curvature term in the Friedmann equation \eqref{eq:friedmann} to the value of the Hubble parameter at the horizon crossing time $t_\ast$ using Eqs.~\eqref{eq:ns-as} and \eqref{eq:a-ast}. For $\omk=-0.005$, and taking the best estimates for $\as$ and $\ns$ from \planck \cite{Ade:2015xua}, we find:
\be
|\omk| \frac{a_0^2H_0^2}{a_\ast^2 H_\ast^2} \simeq 8 \times 10^{-13},
\ee
which shows that the flat approximation is indeed valid at $t_\ast$. In addition, the interval $\ns \pm 2 \sigma_{\ns}$ provided by \planck corresponds to $\epsilon \in (0.006,0.012)$, which is consistent with a slow-roll approximation, showing that assumption (ii) is also valid.

For a given background solution $a(t),\phi(t)$ selected in this manner, we fix
initial conditions for the evolution of the gauge-invariant perturbations
$q_{nlm}(t)$ at the onset of inflation $t_0$. The perturbations are quantized
with the application of standard techniques, and the initial state of the
quantized perturbations is chosen to be the instantaneous vacuum at $t_0$,
defined --- as in the flat case --- as the ground state of the instantaneous
Hamiltonian $H(t_0)$ \cite{Mukhanov:2007zz,Fulling:1989nb}, in the static limit. Such state can be
determined as follows. The equations of motion for the
perturbations, Eq.~\eqref{eq:evolutionQ}, when written for the Mukhanov-Sasaki
variable $v=a~q$ and in terms of the conformal time $\eta$, reduce in the static limit to:
\be
v''_{nlm} + \left[ m^2 a^2 \left( 1+ \frac{6}{n^2-4} \right) + (n^2-1) \right] v_{nlm} = 0 \, .
\ee
Here, the primes represent derivatives with respect to the conformal time $\eta$. These equations describe a set of harmonic oscillators with mode-dependent frequencies
\be
\omega_n = \sqrt{m^2 a^2 \left( 1+ \frac{6}{n^2-4} \right) + (n^2-1)} \, .
\label{eq:mode-frequency}
\ee
Recall that a choice of vacuum state for the
quantized field $q$ is equivalent to a choice of positive frequency solutions $v_{nlm}(\eta)$. In a static FLRW Universe, the vacuum state can be uniquely selected by symmetry requirements: there is a unique regular state invariant under all spatial symmetries and time-translations. Such state is the tensor product of the vacua of each of the normal modes, and is described by positive frequency solutions satisfying:
\be
v_{nlm}(t_0) = \frac{1}{\sqrt{2 \omega_n}}  \, , \qquad v'_{nlm}(t_0) = - i \sqrt{\frac{\omega_n}{2}} \, .
\ee
These conditions correspond to:
\ba
q_{nlm}(t_0) & = & \frac{1}{a_0\sqrt{2\omega_n}} \, , \nonumber \\
\dot{q}_{nlm}(t_0) & = & -\frac{\dot{a}_0}{a_0^2} \frac{1}{\sqrt{2 \omega_n}}-\frac{i}{a_0^2} \sqrt{\frac{\omega_n}{2}} \, .
\label{eq:inst-vacuum}
\ea
In a dynamical background, the instantaneous vacuum at $t_0$ is defined as the vacuum associated with normal modes satisfying \eqref{eq:inst-vacuum}, with frequencies $\omega_n$ given by Eq.~\eqref{eq:mode-frequency} for the instantaneous scale factor $a(t_0)$.

We evolve the initial conditions \eqref{eq:inst-vacuum} for a family of background solutions $(a(t),\phi(t))$ satisfying the condition \eqref{eq:long-enough-inf}. In order to estimate the wavelength of the largest observable mode $\lambda_{\rm long}$, we set $\lambda_{\rm long} = d_{\rm CMB}$, where $d_{\rm CMB}$ is the comoving diameter of the CMB sphere, yielding $k_{\rm long} = 2.3 \times 10^{-4} \textrm{ Mpc}^{-1}$. From  Eq.~\eqref{eq:n-k}, this corresponds to $n_{\rm long} \sim 14$. For such solutions, all observable modes are within the comoving Hubble horizon at the onset of inflation $t_0$. In \Fref{fig:horizons-onset}, we show the evolution of the comoving Hubble horizon around the onset of inflation for two distinct backgrounds and compare it with $1/k_{\rm long}$. The solution `${\rm cond_1}$' represents the case for which the initial conditions for the background geometry are provided using the best fit values of $A_s$ and $n_s$. For this case the largest observable mode is well within the horizon at $t_0$.  The variation of the frequency term in the full Mukhanov-Sasaki equation is then checked to be negligible around $t_0$, with $H \dot{w}_n/\omega_n\ll 1$ for all observable modes $n>14$. The solution `${\rm cond_2}$' represents the extreme case considered in our analysis. In this case, $1/k_{\rm long}$ is close enough to the horizon so that the fractional variation of the frequency becomes of the order of $10\%$ during the few tenths of an e-fold before $t_0$ while $k_{\rm long}$ is within the horizon. 
{This extreme case marks the point in the parameter space for which the static approximation starts becoming unreliable. In the next section, these two conditions illustrate the range of possible effects on the primordial power spectrum and observed anisotropies in the CMB  due to spatial curvature when $\omk=-0.005$.}

\bfig
 \begin{center}
 \includegraphics[width=0.55\textwidth]{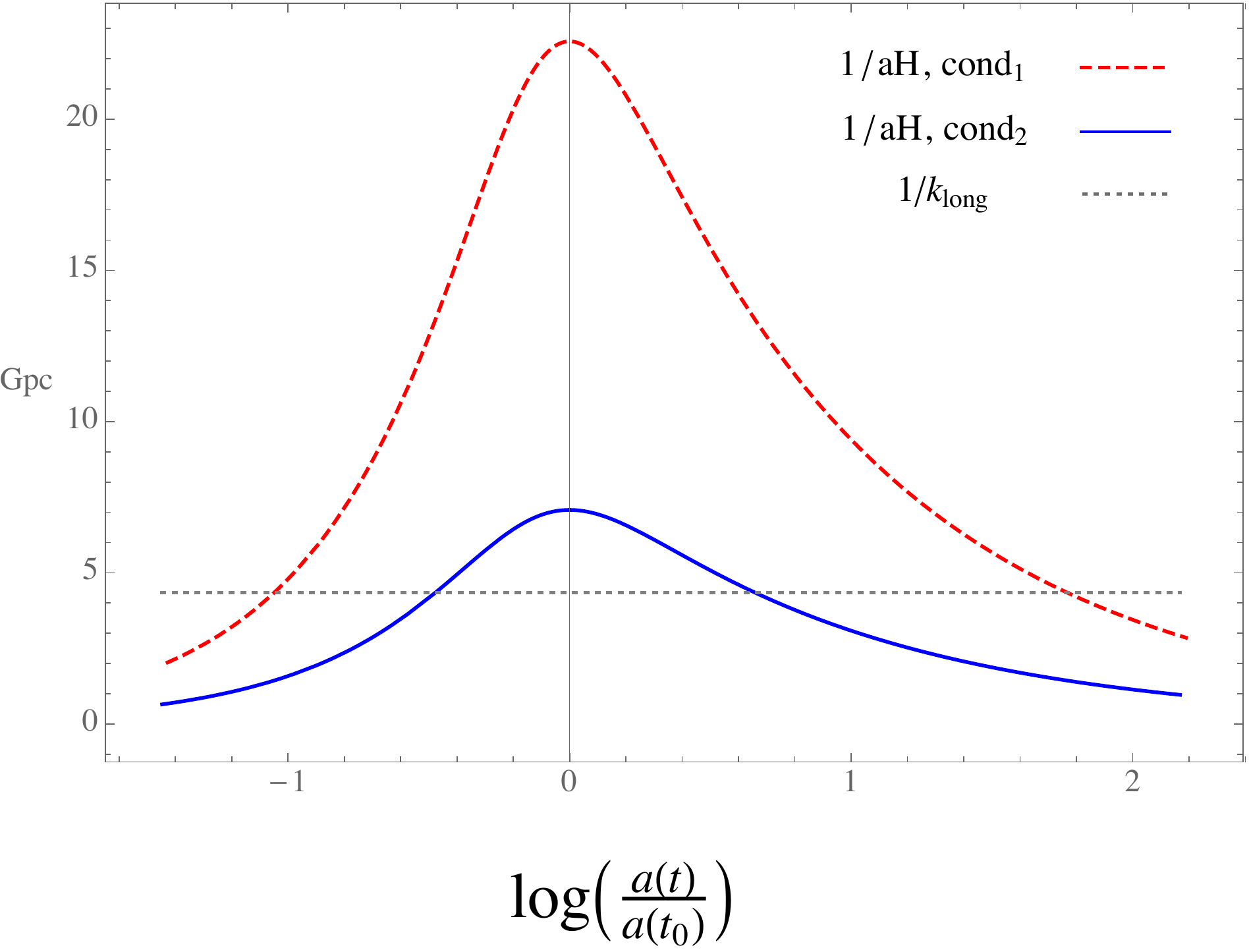}
 \end{center}
 \caption{Comoving Hubble horizon near the onset of inflation for $\Omega_k=-0.005$ with two different sets of initial conditions in the space of allowed initial data. The dashed (red) curve represents a typical solution, with the the largest observable mode $k_{\rm long}$ well within the horizon at the onset of inflation $t_0$.  The solid (blue) curve represents a limiting case, with $1/k_{\rm long}$ close to the comoving Hubble horizon at $t_0$, for which the effects of the spatial curvature are maximal.}
 \label{fig:horizons-onset}
\efig

The case of the Starobinsky potential is treated similarly. The mass parameter $M$ 
in the potential \eqref{eq:starobinsky} is fixed by taking a flat slow-roll 
approximation at horizon crossing for the mode $k_\ast$ and using the number of $e$-folds ($N_*$) between the horizon crossing and the end of inflation. This leads to 
formulas analogous to eqs.~\eqref{eq:a-ast} and \eqref{eq:bg-initial-cond-m} which give (see \cite{bg2} for 
details of estimation of parameters for the Starobinsky potential): 
\ba
\phi_\ast=1.08 \pm 0.02 ~\mpl, \qquad \phid_\ast = (-4.80 \pm
0.70)\times10^{-9}\mpl^2, \\
a_\ast = 2.11\pm0.14 \times 10^{-53}r_o,  \qquad M= (2.56 \pm 0.18) \times 10^{-6}~\mpl.
\label{eq:staroini}
\ea
The initial conditions for the perturbations are also set at the onset of inflation by choosing the instantaneous vacuum determined by the conditions \eqref{eq:inst-vacuum}.

\section{Results}
\label{sec:results}
Given the initial data for the background and the gauge-invariant perturbations, 
we numerically evolve the equations of motion of the background and the quantum
perturbations. This section is divided into two subsections. In the first, we
discuss the evolution of the background spacetime and show that the presence of
spatial curvature can affect the inflationary dynamics and potentially leave
observable imprints on the long wavelength modes. The second subsection describes
the evolution of scalar perturbations, their power spectrum at the end of
inflation and the resulting temperature anisotropy spectrum $C_{\ell}$ observed
in the CMB.

\subsection{Why does spatial curvature matter?}
One of the attractive features of the inflationary paradigm is that it
solves the flatness problem. Starting from a generic initial condition inflation
dilutes away all spatial curvature effects that might have been present before
the onset on inflation. As result at the end of inflation and today the Universe
is extremely close to being spatially flat: $|\omk|<0.005$, i.e., the
contribution of spatial curvature to the total energy density of our Universe today is less than $0.5\%$. 
What about in the past? The post inflationary evolution of 
the Universe is described by the \lcdm model and
the associated Friedmann equation is given by \eref{eq:friedlcdm}. The fraction
of energy density due to the spatial curvature decreases monotonically towards the 
past. Using the current estimates of the cosmological parameters $\omm,~\omr,~\oml$ 
and $\omk$ from {\it Planck}, it is evident from \eref{eq:friedlcdm} that at the 
beginning of the radiation dominated era (approximately 66 \efolds before today), 
which --- assuming instantaneous reheating --- is the same as the end of inflation:
\be
 \left.\f{\rhok}{\rhor}\right|_{\rm beg~rad} \approx 10^{-57} \approx
\left.\f{\rhok}{\rhophi}\right|_{\rm end~inf},  
\ee
where $\rhok=\f{3}{8\pi G}\f{\omk H_0^2}{a^{2}/a_0^2}$, $\rhor=\f{3}{8\pi
G}\f{\omr H_0^2}{a^{4}/a_0^4}$ and $\rhophi=\phid^2/2+V(\phi)$ is the energy density of the inflaton. 
This shows that the contribution of spatial
curvature at the beginning of the radiation era is $10^{-55}\%$. Hence, the effect of 
spatial curvature has been practically insignificant starting from the end of 
inflation till today.

During inflation, however, the spacetime is quasi-de Sitter and $\rhophi$ 
remains nearly constant while $\rhok$ scales as $a^{-2}$.
In the backward evolution the spatial curvature terms keep getting
stronger and approximately $~60$ \efolds before the end of inflation $\rhok$ can
become comparable to $\rhophi$, potentially leading to the breakdown of the
slow-roll phase \cite{Ellis:2001ym,Ellis:2001yn,Uzan:2003nk}. Therefore, the long 
wavelength modes which exit the Hubble horizon during the early stages of 
inflation will be influenced by the spatial curvature and their power spectrum can 
carry the imprints of spatial curvature. This is in contrast to the short 
wavelength modes which exit the Hubble horizon later and thus remain unaffected by 
the spatial curvature. Consequently, their power spectrum is practically the same 
as in the absence of spatial curvature.

\Fref{fig:horizons} shows the evolution of the Hubble horizon, $(aH)^{-1}$,
during inflation and radiation era all the way till the CMB for $\omk<0$
and $\omk=0$. It is evident from the figure that the Hubble horizons for flat and 
closed models differ in the early stages of inflation, while agreeing with each other
extremely well in the future. This indicates that the long wavelength modes 
(denoted by $k_{\rm long}$) which exit the horizon during the early stages of 
inflation may carry imprints of the spatial curvature, while the short 
wavelength modes (denoted by $k_{\rm short}$) which exit the horizon later will 
remain unaffected by the spatial curvature.
\bfig
 \begin{center}
 \includegraphics[width=0.7\textwidth]{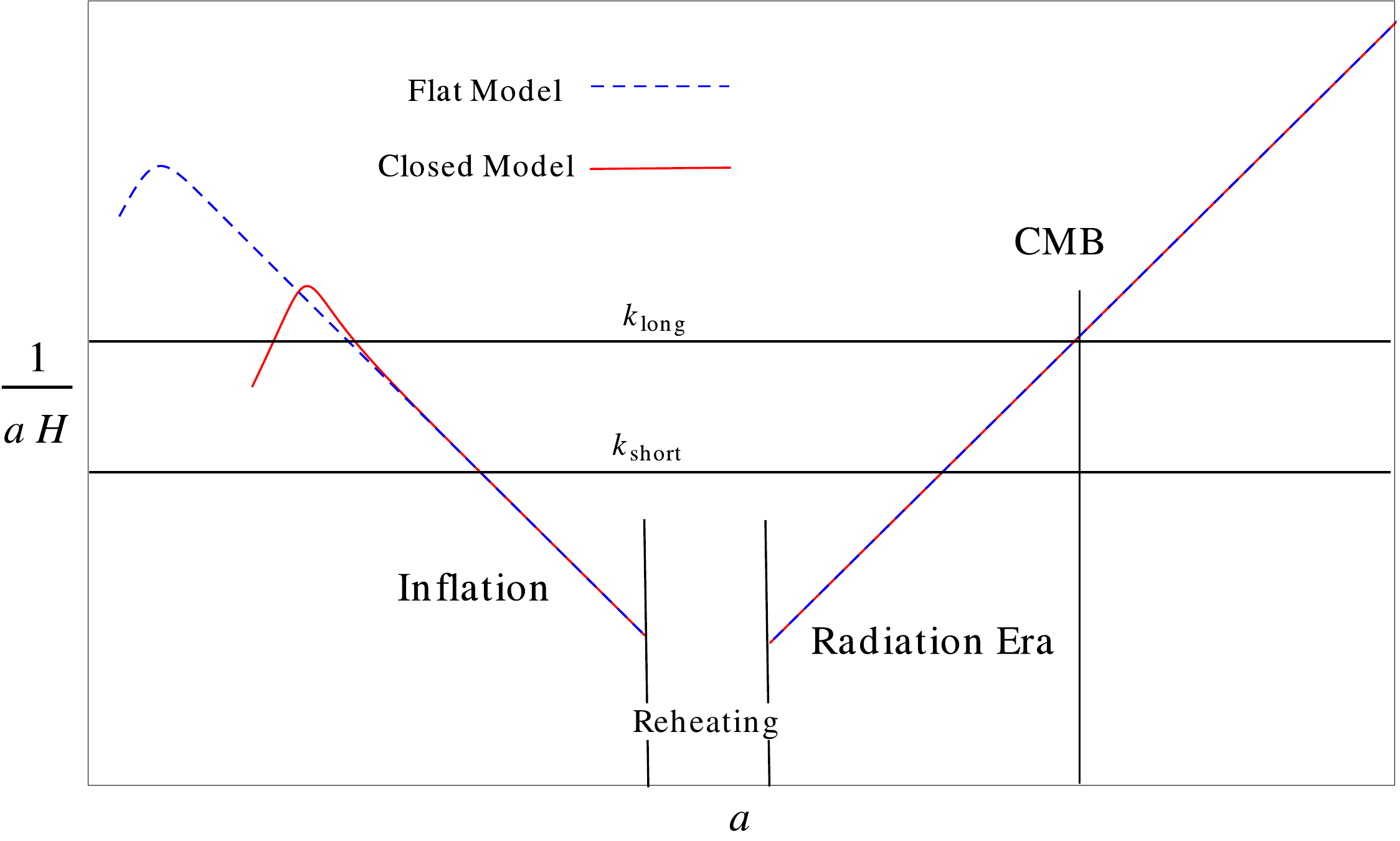}
 \end{center}
 \caption{Comparison of Hubble horizon during and after inflation in a flat FRLW
model and a closed model with $\omk=-0.1$. The difference due to spatial
curvature is important close to the onset of inflation and remains negligible
throughout the rest of the future evolution. This indicates that the long
wavelength modes which exit the curvature radius earlier during inflation may
carry imprints of the spatial curvature. These potential imprints will, for instance, affect their power spectrum. }
 \label{fig:horizons}
\efig

{\bf Remark:} Note that it is not {\it a priori} clear whether or not 
inflation will last long enough in the presence of spatial curvature. As 
pointed out in \cite{Uzan:2003nk}, the amount of inflation in the closed model 
is bounded above and the bound is dictated by the parameter $\omk$. As a 
result, the space of initial conditions leading to sufficient amount of 
\efolds in the closed model will be smaller than for the flat model. In 
this paper, we are interested in finding at least one set of initial conditions for 
which the desired inflationary phase takes place. As described in the previous 
section, we found a family of initial conditions on $\phi$, $\phid$ and $a$ 
which lead to the desired inflationary phase and are compatible with observations 
within the error bars. A more quantitative analysis of the space of initial 
conditions and the viability of desired inflationary phase requires 
investigation of a suitable measure on the space of initial data and will be 
studied in detail in a quantum gravitational extension of this paradigm in \cite{bgy3}. 

Let us now study the evolution of the scalar perturbations and analyze the
potentially observable imprints of spatial curvature on the CMB.

\subsection{Power spectrum}
\bfig
 \ig[width=0.5\textwidth]{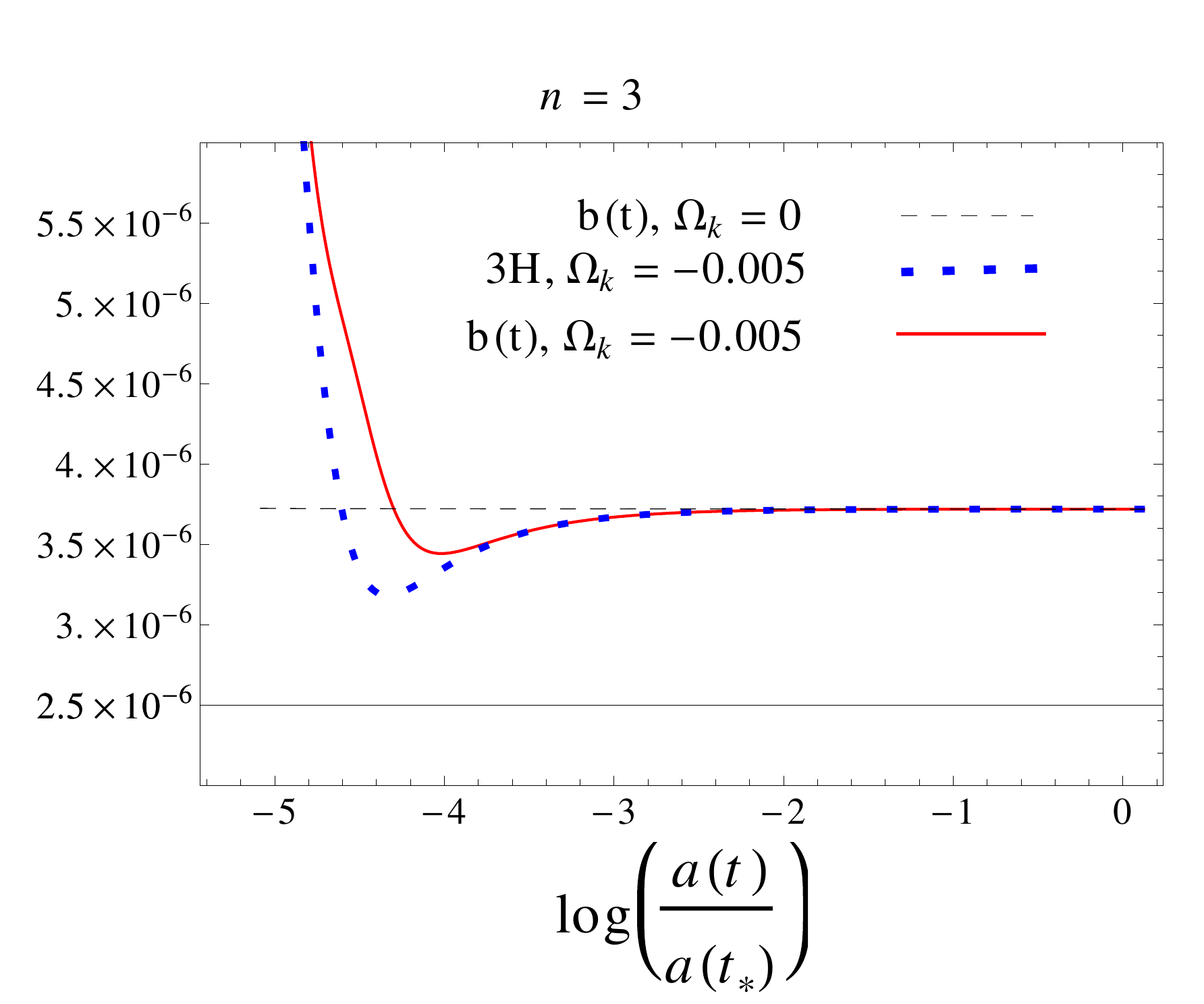}
 \ig[width=0.5\textwidth]{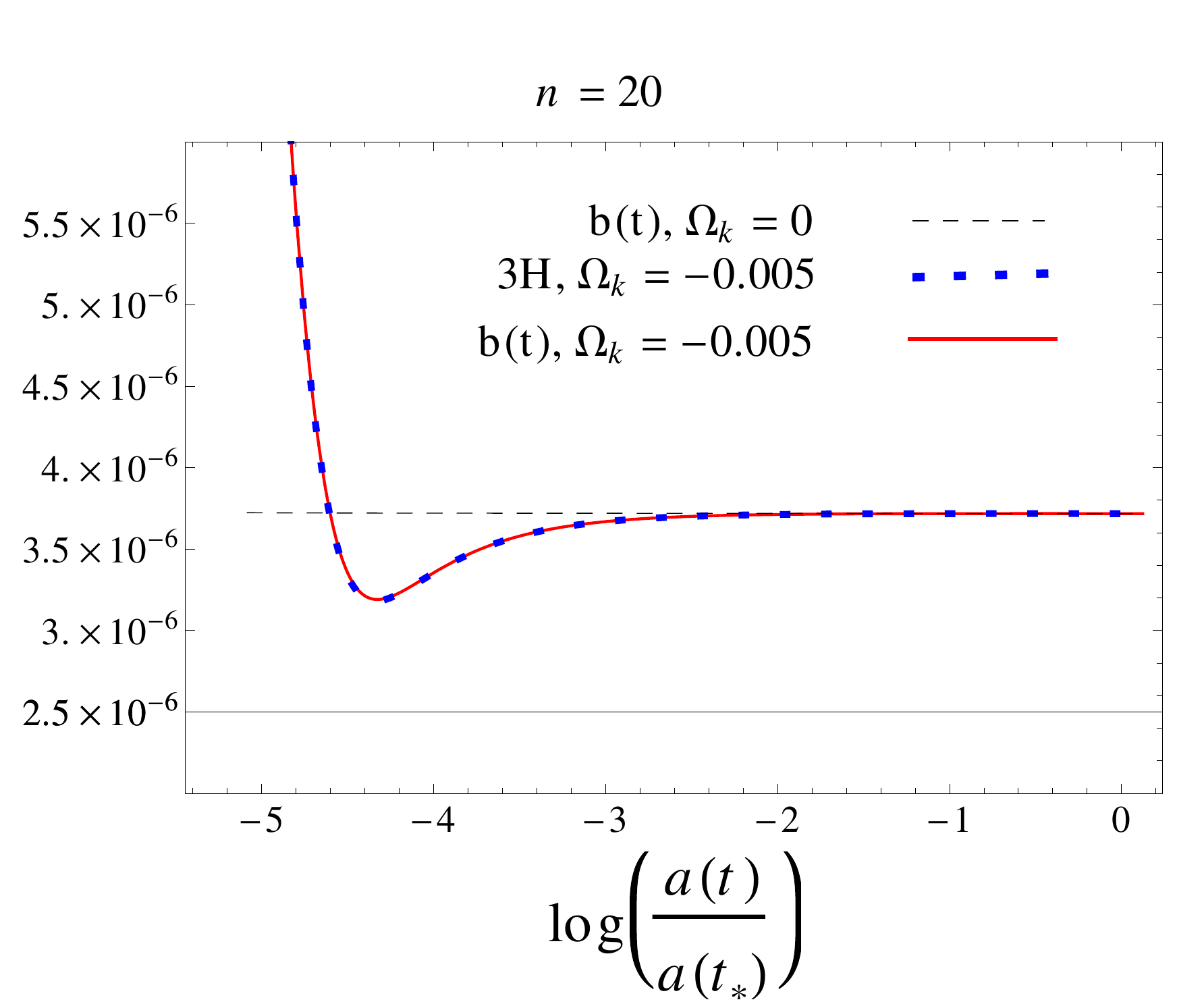}
\caption{Comparison of the coefficient $b(n,t)$ in the evolution equation 
(\ref{eq:evolutionQ}) for the gauge-invariant scalar perturbation $\Q_{nlm}$ in 
a closed FLRW model for $\omk=-0.005$ with $3H$ in the same closed model and with 
$b(n,t)$ for a flat model. The left panel corresponds to the wavenumber $n=3$ for 
which $b(n,t)$ is distinct from $3H$ in the closed model. The right panel corresponds
to $n=20$ for which the two quantities are practically indistinguishable. 
In both cases, $b(n,t)$ for the flat model is different from $3H$ and $b(n,t)$ for the 
closed model. These differences become more prominent for larger magnitudes of 
$\omk$.}
\label{fig:bcoeff}
\efig
\bfig
 \ig[width=0.5\textwidth]{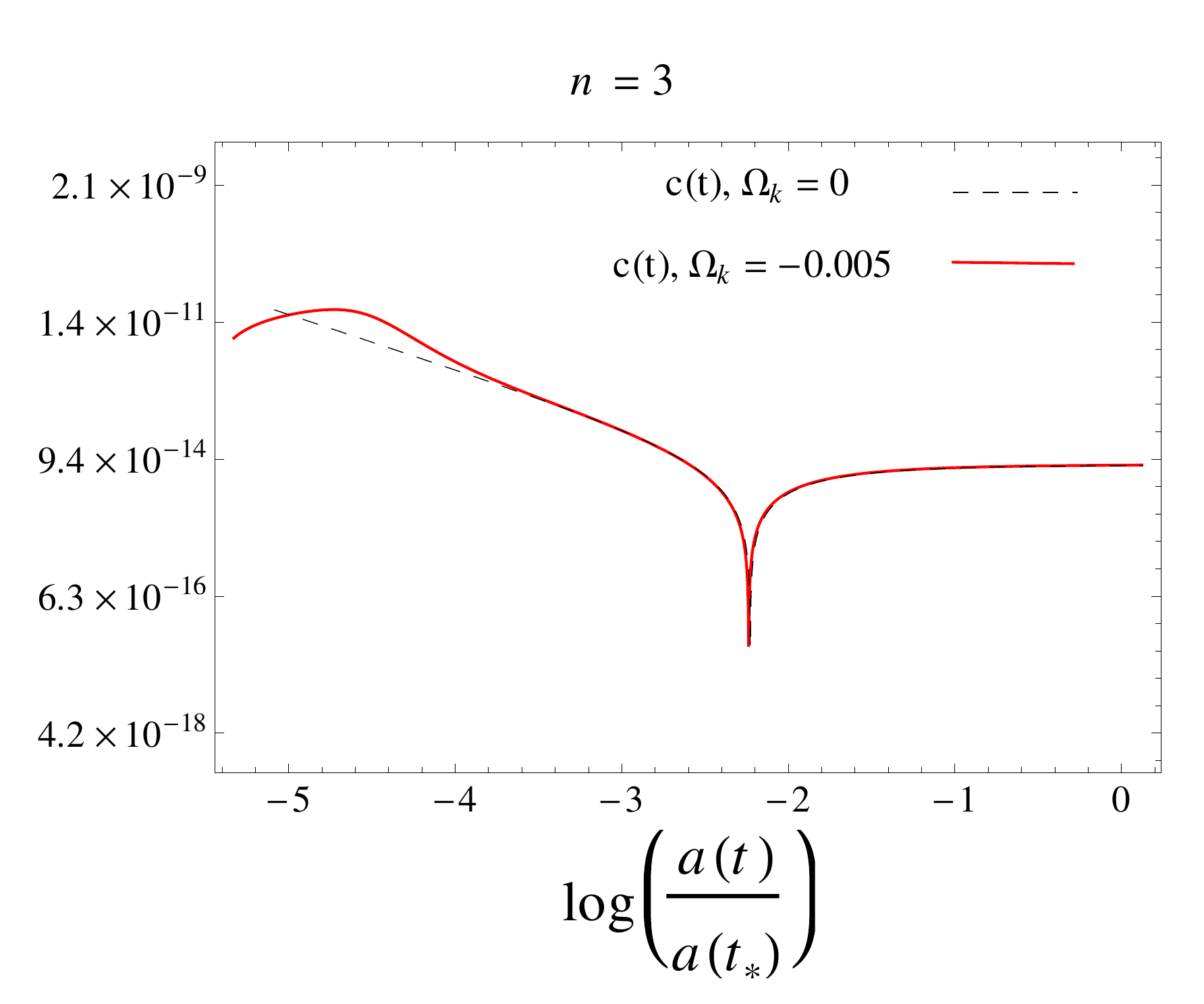}
 \ig[width=0.5\textwidth]{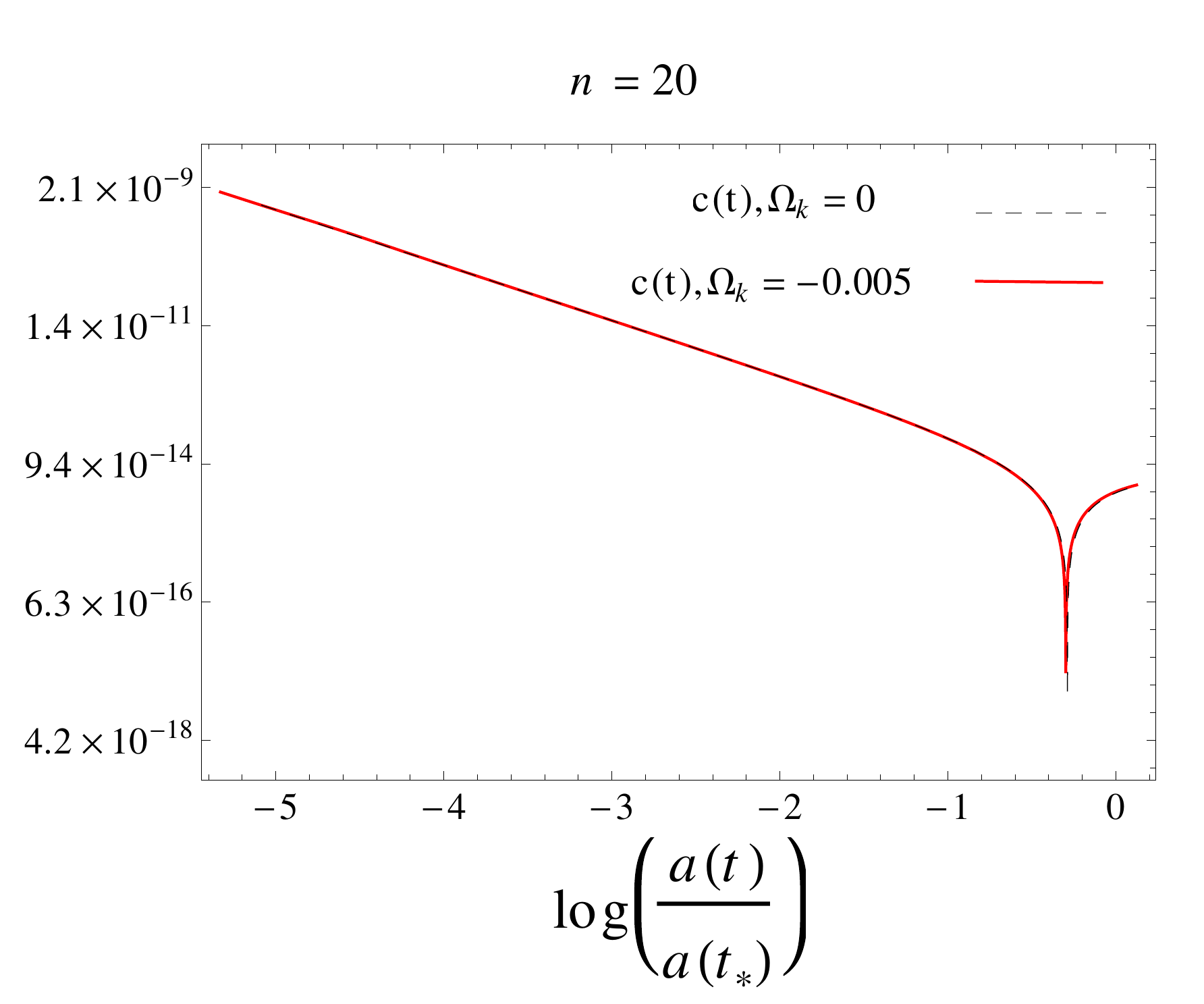}
\caption{Comparison of the coefficient $c(n,t)$ in the evolution equation 
(\ref{eq:evolutionQ}) for the gauge-invariant scalar perturbation $\Q_{nlm}$ in 
a closed FLRW model for $\omk=-0.005$ with that in a flat model. The left panel 
corresponds to the wavenumber $n=3$ for which $c(n,t)$ in closed model shows 
deviations from that in the flat model. The right panel corresponds
to $n=20$ for which the two curves are practically indistinguishable.}
\label{fig:ccoeff}
\efig

We now consider the evolution of the gauge-invariant quantum scalar
perturbations $\Q_{nlm}$ on the inflationary background geometry discussed above.  
As discussed in \sref{sec:inidata}, $b(n,t)$ and $c(n,t)$, the coefficients of 
$\dot q$ and $q$ in the equation of motion for $\Q$ in the 
closed model (see \eref{eq:evolutionQ}),
take a very different form compared to the flat FLRW model. In particular, 
$b(n,t)$ for the flat model is simply $3 H(t)$, whereas in the closed model  
$b(n,t)$ also depends on the comoving wavenumber $n$ in a rather complicated fashion. 
Therefore, the evolution of scalar perturbations in a closed inflationary model 
is different from the flat model in two ways. First, the evolution of 
background quantities $H(t)$ and $a(t)$ are different which leads to differences 
in the causal horizon (\fref{fig:horizons}) and second, the evolution equation 
for the perturbations acquires additional $n$ dependence for the 
closed model whose numerical effect becomes more important for small $n$. 
\Fref{fig:bcoeff} and \ref{fig:ccoeff} respectively show the evolution of 
the coefficients $b(n,t)$ and $c(n,t)$ in \eref{eq:evolutionQ} for the closed model 
compared with that in the flat model during inflation. The horizontal axes in both 
plots show the number of \efolds from the time $t_*$ when the reference mode 
$k_*=0.002~\mpc$ exits the Hubble horizon. It is apparent from the figures that 
approximately 4 \efolds before the horizon exit:
\begin{itemize}
\item $b(n,t)$ for the closed model is different from that in the flat model for both 
$n=3$ and $n=20$ (\fref{fig:bcoeff}),
\item for large $n$, $b(n,t)$ in the closed model approaches $3H$ (right panel of
\fref{fig:bcoeff}) and 
\item $c(n,t)$ in the closed model shows deviations from that in the flat model for 
small $n$ (e.g., $n=3$ shown in the left panel of \fref{fig:ccoeff}) while for 
large $n$ they are practically indistinguishable (e.g., $n=20$ shown in the
right panel of \fref{fig:ccoeff}). 
\end{itemize}
Therefore, the spatial curvature effects are more prominent during the early stages of 
inflation and for small $n$ modes which exit the Hubble horizon at these times.

\bfig
 \begin{center}
  \ig[width=0.6\textwidth]{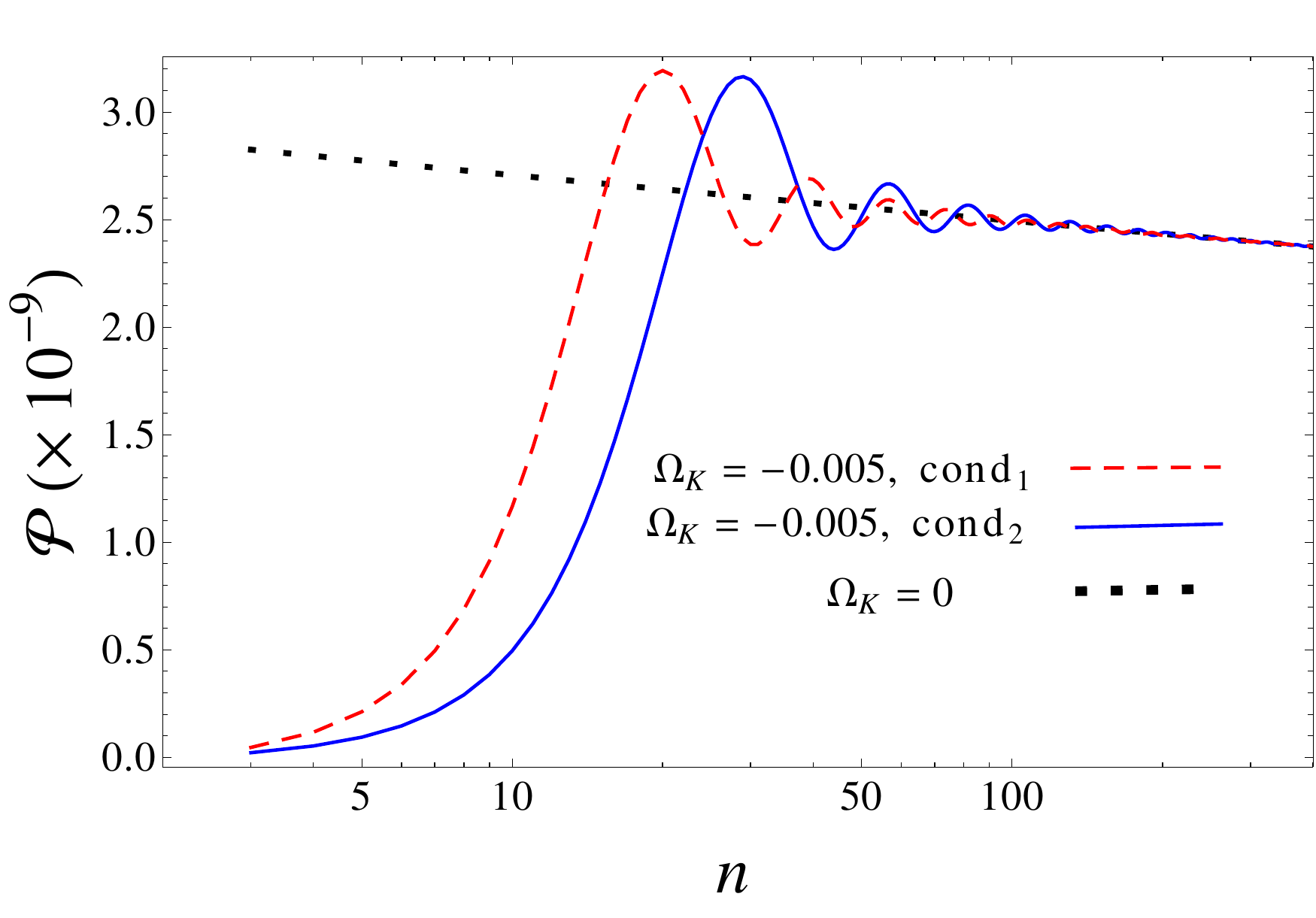}
 \end{center}
  \caption{Scalar power spectra for the closed model for $\omk=-0.005$ with two
different sets of initial conditions for the background geometry. The dashed (red) 
curve corresponds to: $\phi(t_*)= 3.07~\mpl$,
$\phid(t_*)=-2.08\times10^{-7}~\mpl^2$ and the solid (blue) curve corresponds to: 
$\phi(t_*)=2.99~\mpl$, $\phid(t_*)=-2.13\times10^{-7}~\mpl^2$. The dotted straight 
curve shows the almost scale-invariant spectrum. The closed model 
power spectrum shows oscillations for small $n$ and approaches the flat power  
spectrum for large $n$. Interestingly, the average power in the small 
$n$ regime is suppressed compared to the usual nearly scale-invariant power
spectrum. Here we have shown the power spectrum for $3\leq n\leq400$. However, 
for $\omk=-0.005$, only the modes with $n\gtrsim14$ are observable.}
  \label{fig:powerspec}
\efig

Recall that in the flat model the initial conditions for the perturbations are given 
a few \efolds before the mode with the longest wavelength exits the curvature radius 
and the scalar perturbations are assumed to be in a Bunch-Davies vacuum state as the 
background geometry behaves like a quasi-de Sitter spacetime.  
In contrast, for the closed model, the spacetime shows deviations from a quasi-de 
Sitter spacetime at the onset of inflation as shown in \fref{fig:horizons}. 
Consequently, the Bunch-Davies approximation is violated. However, since all the 
observable modes are still inside the curvature radius, we provide the initial 
conditions assuming static initial conditions as described in \sref{sec:inidata}. 

We explore all observationally compatible initial conditions for the background 
geometry for the two potentials considered: the quadratic potential 
(see \eref{eq:quadini}) and Starobinsky potential (see \eqref{eq:staroini}). 
{In \fref{fig:powerspec} we show scalar power spectra for  the two initial conditions 
discussed in section \ref{sec:inidata} for the quadratic potential with 
$\omk=-0.005$. The initial conditions are chosen such that:
(i) `${\rm cond_1}$' (red, dashed curve) for which $\phi(t_*)= 3.07~\mpl$, $\phid(t_*)=-2.08\times10^{-7}~\mpl^2$
at the time $t_*$ when the reference scale $\ks=0.002~\mpc$ exits 
the curvature radius, and 
(ii) `${\rm cond_2}$' (blue, solid curve) for which $\phi(t_*)=2.99~\mpl$, $\phid(t_*)=-2.13\times10^{-7}~\mpl^2$ at
the same time $t_*$.
} The usual nearly scale-invariant spectrum is shown by the black dashed straight line. The power spectra for 
the closed model have two main features. First, the closed model power spectrum is 
the same as that in the flat model for large $n$. These modes correspond to short 
wavelength modes that exit the Hubble horizon later during inflation. 
Second, for small $n$ (corresponding to longer wavelength modes which exit the 
Hubble horizon earlier during inflation) the closed model power spectrum 
oscillates around the flat one. Interestingly, these oscillations when averaged 
over a range of $n$ show that: $(\Pclosed)_{\rm avg}<(\Pflat)_{\rm avg}$. That
is, the closed model power spectrum has a power deficit at large scales
compared to the flat model. Numerical computations with the Starobinsky
potential give rise to similar results. Note that this may not be true for the 
tensor perturbations, because heuristically similarly to the flat case where the 
tensor-to-scalar ratio is
different for the two potentials while the scalar power spectrum is the same
\cite{Ade:2015lrj}. 

The question now is: are these features strong enough to have observational 
imprints on the temperature anisotropy spectrum observed in the CMB? For 
instance, do these oscillations show up in the temperature anisotropy spectrum of
the CMB ($\celltt$) observed today and more importantly, 
does the deficit of power in the power spectrum at the end of inflation lead 
to a suppression of power in the CMB, which is one of the large scale anomalies 
reported by the recent \planck and WMAP missions? 

\subsection{Temperature anisotropy spectrum}
\bfig
 \begin{center}
  \ig[width=0.8\textwidth]{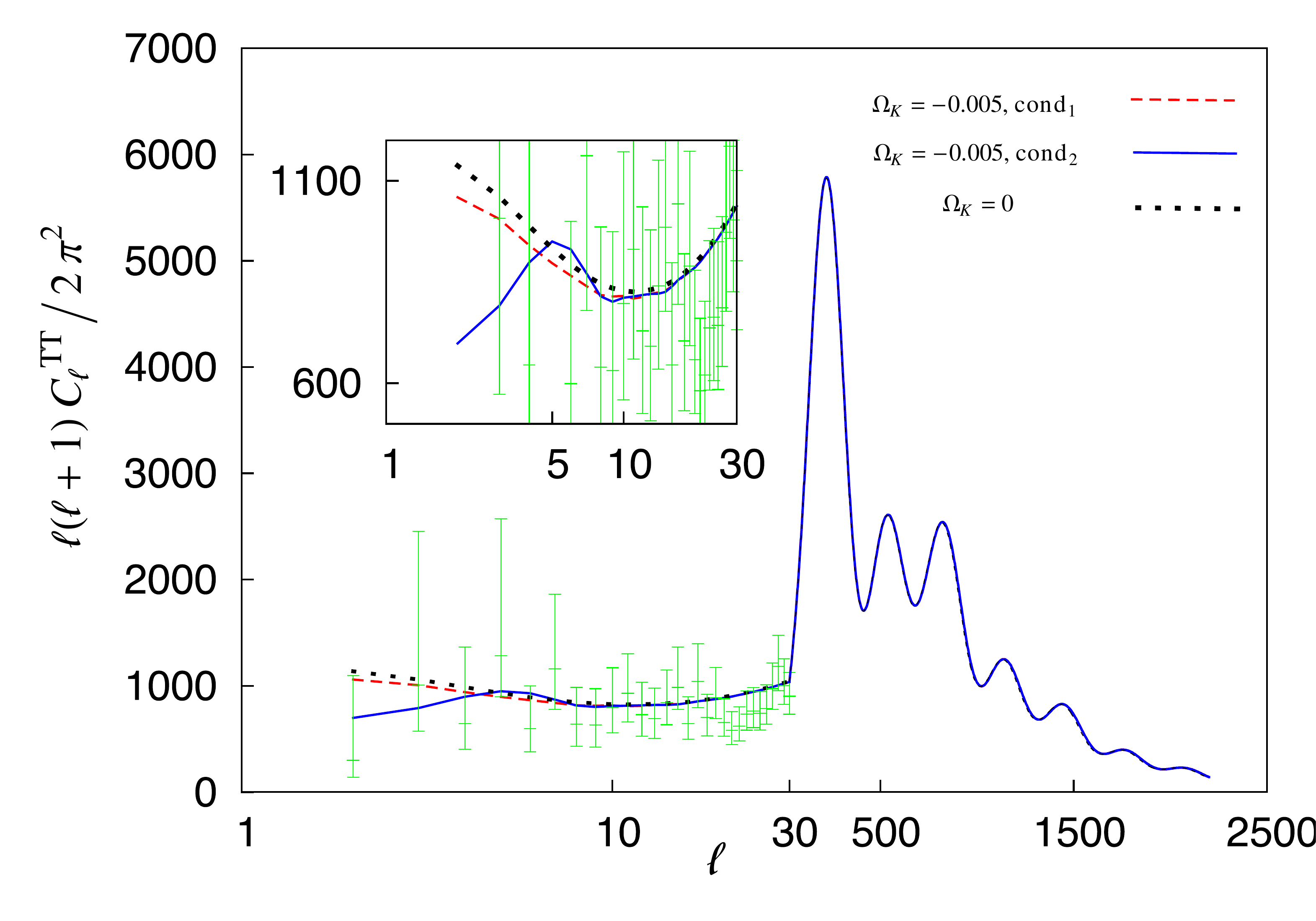}
 \end{center}
  \caption{The temperature anisotropy power spectrum for closed (dashed and
           solid curves) 
           and flat (dotted) models and the {\it Planck} 2015 data with error  
           bars. The dashed (red) curve corresponds to: $\phi(t_*)= 3.07~\mpl$, $\phid(t_*)=-2.08\times10^{-7}~\mpl^2$. The solid (blue) curve corresponds to : 
           $\phi(t_*)=2.99~\mpl$,
$\phid(t_*)=-2.13\times10^{-7}~\mpl^2$, both with $\omk=-0.005$. 
           It is evident that in both cases the closed model shows deficit 
           power at low $\ell$'s compared to the flat model at $\ell\lesssim10$.}
  \label{fig:cls}
\efig
To answer these questions, we evolve the scalar power spectrum at the end of
inflation using the Boltzmann code \camb to compute the temperature anisotropy
spectrum $\celltt$ at the surface of last scattering \cite{Lewis:1999bs}. 
\Fref{fig:cls} shows the resulting temperature anisotropy power spectrum  
for a closed model with $\omk=-0.005$ with initial conditions corresponding to
the power spectra in \fref{fig:powerspec}. The dashed (red) curve corresponds to 
$\phi(t_*)= 3.07~\mpl$, $\phid(t_*)=-2.08\times10^{-7}~\mpl^2$ 
(denoted as ${\rm cond}_1$ in the figure) and the  solid (blue) one corresponds 
to $\phi(t_*)=2.99~\mpl$, $\phid(t_*)=-2.13\times10^{-7}~\mpl^2$ (denoted as
${\rm cond}_2$ in the figure). It is evident from the figure that the closed model 
$C_\ell$ agrees extremely well with the {\it Planck} data and the predictions of the 
standard inflationary scenario with flat FLRW model at large $\ell$.
Interestingly, for $\ell\lesssim 20$, however, the closed model shows deficit of 
power as compared to the flat model as expected from \fref{fig:powerspec}. The 
second, more extreme condition has less number of \efolds during inflation compared 
to the first condition, leading to suppression as large as $\sim30\%$ in $\celltt$ 
at $\ell=2$, while the first condition leads to $\sim10\%$ suppression there. 
The corresponding E-mode 
polarization spectrum and its cross correlation with the temperature spectrum are 
shown in \fref{fig:clsPol}, which show small deviations from the flat model at 
similar scales. The suppression in power is limited to the multipoles $\ell<20$, 
and therefore is not enough to explain the power suppression anomaly observed in 
the CMB. Nevertheless, as is evident from our analysis, it can contribute to the
suppression of power at these scales. 
\bfig
  \ig[width=0.5\textwidth]{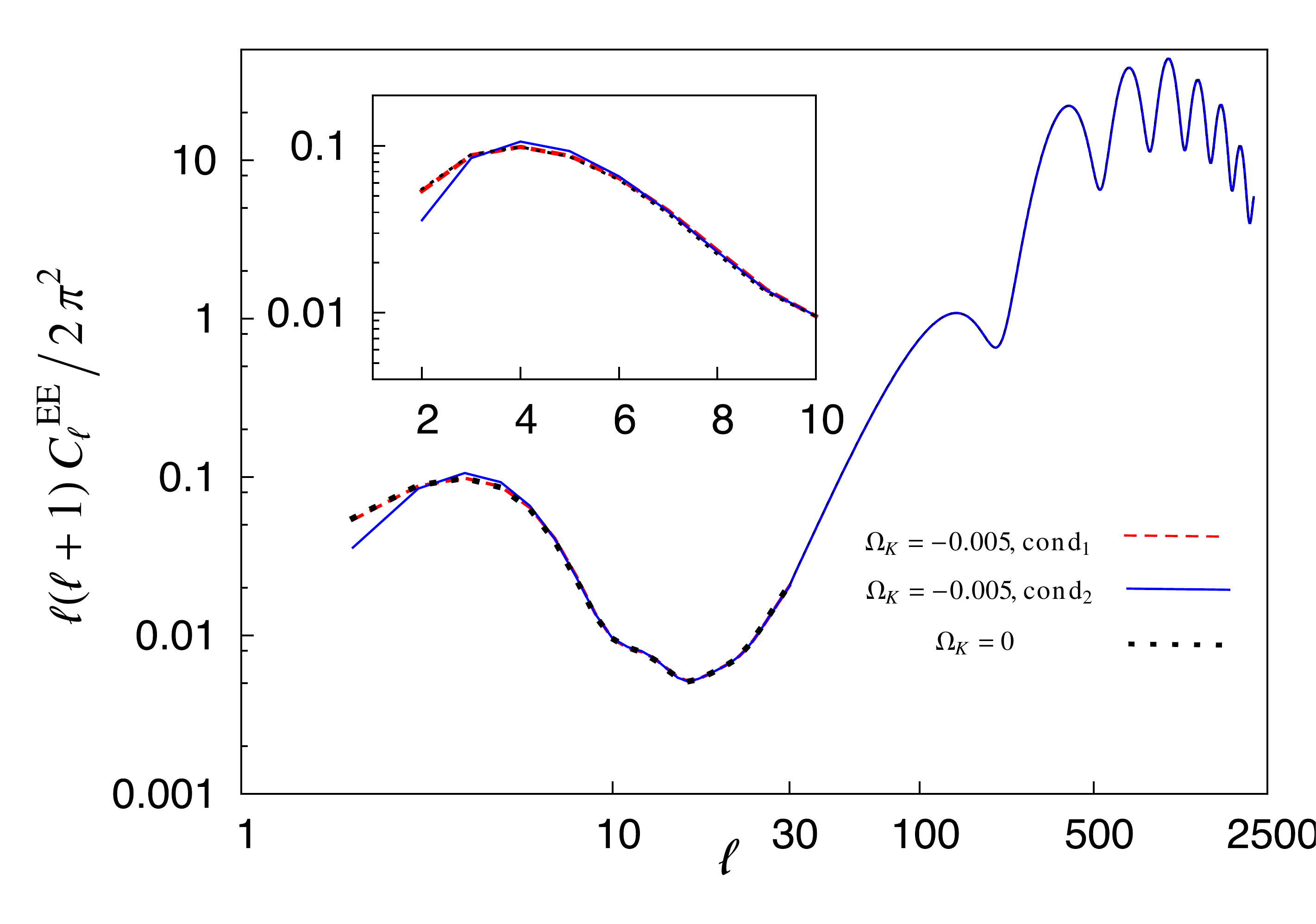}
  \ig[width=0.5\textwidth]{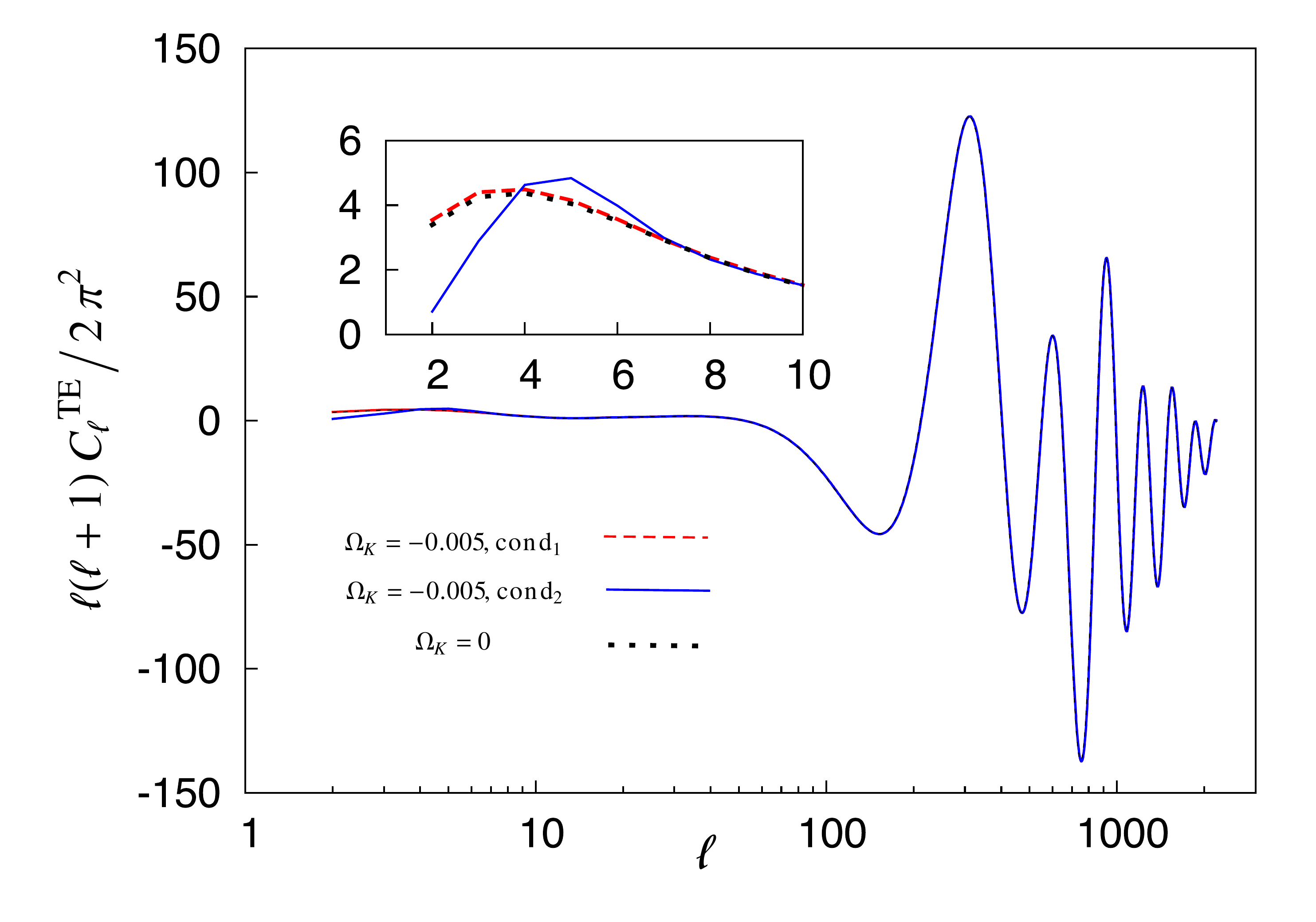}
  \caption{E-mode polarization spectrum and cross correlation with the temperature
anisotropy spectrum for the closed model with the same initial conditions as in
\fref{fig:cls}. There are tiny differences between the closed model and the flat
model polarization spectrum at $\ell<10$.}
  \label{fig:clsPol}
\efig

Here for the concreteness of the discussion 
we have shown the plots for $\omk=-0.005$ which is the bound on spatial curvature 
given by combining \planck and BAO data. {If one takes a higher magnitude of $\omk$ 
the power suppression will be stronger, and conversely, smaller $|\omk|$ will show less power suppression.}
In order to find the best fit value of $\omk$ 
using the modified primordial power spectrum obtained here, we revisit the the estimation of 
cosmological parameters by comparing the resulting $\celltt$ against the 
recent \planck and BAO data using the publicly available Markov Chain Monte Carlo 
code \cosmomc \cite{Lewis:2002ah}. We allow variations of all cosmological 
parameters. We find that, since the modifications are limited to very low $\ell$'s, 
they practically have no effect on the parameter estimation. In particular, we 
found: $\omk=0.000\pm0.005$, which is exactly what is reported in 
\cite{Ade:2015xua}. Therefore, although conceptually the \planck estimation of 
$\omk$ in \cite{Ade:2015xua} is inconsistent, the estimation of cosmological 
parameters in the presence of spatial curvature is phenomenologically robust. 

\section{Discussion}
\label{sec:disc}
The presence of spatial curvature can affect the CMB observations in two ways: 
modifications in the Boltzmann equations which affect the transfer function from 
the end of inflation till today and in the spectrum of primordial fluctuations 
at the end of inflation. The former possibility has been studied (see e.g. 
\cite{Lewis:1999bs}), 
while the latter possibility has only been addressed using approximate methods
\cite{Lasenby:2003ur,Efstathiou:2003hk,Masso:2006gv}. However, a full treatment of 
the primordial fluctuations during inflation in the presence of positive spatial 
curvature had been missing so far. In this paper, we studied the inflationary 
dynamics of the closed FLRW model in the presence of the quadratic and Starobinsky 
potential and analyzed the evolution of the cosmological perturbations. We
first obtained the gauge-invariant scalar perturbations following Hamiltonian
methods described in \cite{Langlois1994} and derived the quadratic Hamiltonian that
governs their evolution on the inflationary background geometry. These
perturbations were then numerically evolved by providing suitable initial
conditions before the onset of slow-roll. We computed the resulting power spectrum 
at the end of inflation, which provided the initial conditions for the 
Boltzmann equations that describe the evolution of the cosmological linear
perturbations in the post inflationary phase. We then computed the temperature 
anisotropy spectrum at the surface of last scattering using {\texttt{CAMB}} \cite{Lewis:1999bs}. Naturally, a modification in the primordial power spectrum can leave
observational imprints in the temperature and polarization spectrum of the CMB
and can alter the estimation of cosmological parameters. 

Our analysis shows that not only the background inflationary dynamics of a closed 
FLRW model is different from that in the flat model, but also the scalar power
spectrum at the end of inflation carries signatures of these differences. 
This further leads to observational imprints on the long wavelength modes observed 
in the CMB. We performed numerical computation of $\celltt$ by exploring initial 
conditions for the background geometry allowed by the observational error bars. 
We found that the scalar power spectrum in closed model shows 
oscillatory behavior for long wavelength modes whose wavelengths are comparable to 
the size of the observable Universe today, a feature not present in the flat model. 
For short wavelength modes, on the other hand, there is excellent agreement 
with the predictions of the spatially flat inflationary spacetime. 
Interestingly, we find that although the oscillatory behavior of the power 
spectrum is diluted in the temperature power spectrum for short scales
(see \fref{fig:cls}), there is deficit of power at low $\ell$'s compared to 
the predictions of the spatially flat model. For $\omk=-0.005$, the suppression 
ranges between $~10-30\%$ for $\ell=2$ depending on the initial conditions. In the 
limit $\omk\rightarrow0$ the power spectrum in the flat model, i.e. 
without oscillations and suppression, is 
recovered. Since the suppression is limited to $\ell<10$, it is not 
enough to explain the anomaly observed by recent CMB experiments. Nonetheless, our 
analysis shows that these anomalies can be partially accounted for by the inclusion of 
positive spatial curvature (i.e. with $\omk<0$).

In the estimation of $\omk$ in \cite{Ade:2015xua}, the primordial power spectrum 
was assumed to be nearly scale-invariant. This is conceptually inconsistent, as 
the spatial curvature modifies the primordial power spectrum. 
For $\omk=-0.005$, which is within the \planck constraints, these modifications
are in the observable range. Nevertheless, since the observational 
effects of the spatial curvature are limited to very low multipoles, spatial
curvature has negligible effects on the estimation of the cosmological
parameters. Therefore, estimation of all cosmological parameters, including
$\omk$, by \planck in \cite{Ade:2015xua} is phenomenologically robust under the
inclusion of positive spatial curvature. 

In this paper, we restricted our analysis to initial conditions compatible
with observations. In order to study the naturalness and the origin of these
initial conditions for inflation in the closed FLRW model, one needs to extend the 
analysis to the Planck scale and introduce a suitable measure on the full space
of initial conditions. These questions have 
already been addressed in the setting of loop quantum cosmology (LQC) for the flat 
FLRW model in the presence of quadratic and Starobinsky potential
\cite{Ashtekar:2009mm,Ashtekar:2011rm,Corichi:2010zp,bg2}. In a future work
\cite{bgy3}, we will use the LQC model of the closed FLRW spacetime developed in
\cite{apsv} to address these issues and extend the framework of quantum fields on
quantum spacetime of \cite{Ashtekar:2009mb,Agullo:2012fc} to evolve cosmological
perturbations all the way from the Planck scale to the end of inflation. Another
natural extension of this work is to study the evolution of tensor modes in
closed inflationary FLRW spacetime \cite{bgy2}. A priori, it is not clear whether 
the tensor modes will also show a similar suppression as the quadratic 
Hamiltonian for tensor modes is quite different from that of the scalar modes. 
Spatial curvature effects in the tensor spectrum  may lead to interesting 
features in the B-mode polarization spectrum at large scales which may
become observable in future CMB experiments. In recent works 
\cite{Schmidt:2012ky,Agullo:2015aba,Adhikari:2015yya}, it has 
been shown that coupling between long and short wavelength modes in presence of
non-gaussianity in the flat model can lead to hemispherical asymmetry which is 
another CMB anomaly observed in the CMB. In future work, it will be interesting to study 
how the presence of positive spatial curvature can affect these results.

\vskip0.04cm
\section{Acknowledgements}{We are grateful to Abhay Ashtekar for ample 
discussions, guidance and suggesting this problem. We would also like to thank 
Eugenio Bianchi, Sarah Shandera and Donghui Jeong for discussions.
This work was supported by 
NSF grant PHY-1505411, the Eberly research funds of Penn State and a Frymoyer 
Fellowship to BB. NY acknowledges support from  CNPq, Brazil. This work used the Extreme Science and Engineering Discovery Environment (XSEDE), which is supported by National Science Foundation grant number ACI-1053575.}
\vskip1cm
\appendix
\section{Hyperspherical harmonics and linear perturbations on $\mathbb S^3$}
\label{app:A}

In this appendix we review basic properties of hyperspherical harmonics on the
unit sphere $\mathbb S^3$ and discuss the expansion of linear cosmological perturbations in general closed FLRW Universes in terms of such normal modes (for more details, see \cite{abbott1986,Halliwell:1984eu}). We choose coordinates in which the metric of $S^3$ is of the form:
\be
d\Omega^2 = d\chi^2 + \sin^2 \chi (d\theta^2 + \sin^2 \theta \, d\varphi^2) \, ,
\label{eq:angular}
\ee
with $\chi,\theta \in [0,\pi]$, $\varphi \in [0,2\pi]$. We write the metric tensor as $\Omega_{ij}$, the covariant derivative associated with $\Omega$ as $\nabla$, and the corresponding Laplace-Beltrami operator as $\nabla^2$. 

A complete set of scalar functions on $\mathbb S^3$ is given by the hyperspherical harmonics:
\be
Q_{nlm}(\chi,\theta\,\varphi) = \Phi^l_n(\chi) Y_{lm}(\theta,\varphi) \, ,
\ee
where the $Y_{lm}(\theta,\varphi)$ are spherical harmonics on the $2$-sphere and the Fock harmonics $\Phi_n^l(\chi)$ are defined as:
\be
\Phi^l_n(\chi) = \sqrt{ \frac{M_{nl}}{\sin \chi} } P_{-1/2+n}^{-1/2-l}( \cos \chi) \, , \qquad 
M_{nl} = \Pi_{r=0}^l (n^2 - r^2) \, ,
\ee
where $P_{-1/2+n}^{-1/2-l}$ are the associated Legendre functions.
The hyperspherical harmonics are eigenfunctions of the Laplacian:
\be
\nabla^2 Q_{nlm} = - (n^2-1) Q_{nlm} \, ,
\ee
and satisfy the orthogonality relations:
\be
\int_{\mathbb S^3} d\Omega \, Q_{nlm} Q_{n'l'm'} = \delta_{nn'} \delta_{ll'} \delta_{mm'} \, .  
\ee
We can use the scalar harmonics $Q_{nlm}$ to define a set of orthogonal vector harmonics:
\be
P_i^{nlm} = \f{1}{n^2-1}\nabla_i Q_{nlm}
\ee
which satisfy $\nabla^2 P_i = -(n^2-3) P_i$  and are normalized according to:
\be
\int_{\mathbb S^3} d\Omega \, \Omega^{ij} P_i^{nlm} P_j^{nlm} = \frac{1}{n^2-1}  \, .
\ee
We also introduce the tensor harmonics:
\ba
\Qij &=& \f{1}{3} \Omega_{ij} Q_{nlm} \, , \\
\Pij &=& \left(\f{1}{n^2-1} \nabla_i\nabla_j + \f{1}{3} \Omega_{ij}\right) Q_{nlm} \, ,
\ea
satisfying $\nabla^2 \Qij=-(n^2-1) \Qij$ and $\nabla^2 \Pij = -(n^2-7) \Pij$. These functions are orthogonal for distinct labels $n,l,m$, and satisfy the relations:
\ba
 \int d\Omega~\Qij \Pupij &=& 0 \\
 \int d\Omega~\Qij \Qupij &=& \f{1}{3} \\
 \int d\Omega~\Pij \Pupij &=& \f{2}{3}\f{n^2-4}{n^2-1}\, ,
\ea
where the sum is only over $i$ and $j$ and not over $n$, $l$ and $m$.
The $\Pij$ are traceless and it can be easily verified that:
\be
\nabla^i\nabla^j \Pij = \f{2}{3} (n^2-4) Q_{nlm} \, .
\ee

As in the flat case, cosmological linear perturbations on a closed FLRW Universe can be decomposed into decoupled scalar, vector and tensor modes. For the case of scalar perturbations in the presence of a scalar field, the metric and field perturbations are described by eqs. \eqref{eq:metric-perturb} and \eqref{eq:field-perturb}, while the conjugate momenta are:
\ba
\delta \tilde{\pi}^{ij} &=& \f{1}{a^2 6 \sqrt{V_o}} \sumn
                \left(3 \pi^{(1)}_{nlm} \Qij + \f{3}{2} \f{n^2-1}{n^2-4} \pi^{(2)}_{nlm}
\Pij \right) \sqrt{\Omega} \, , \\
\delta \tilde{\pi}_\phi &=& \sqrt{\f{4\pi G}{3 \V0}} \sumn \pi_\phi^{nlm} Q_{nlm}
\sqrt{\Omega} \, .
\ea
The coefficients in the expansion of the scalar perturbations and conjugate momenta are determined by:
\ba
     \gamma^{(1)}_{nlm} &=& \f{1}{2 a^2 \sqrt{V_o}} 
                \int d\Omega~\dgamma_{ij} \Qij \, , \\
     \gamma^{(2)}_{nlm} &=& \f{1}{4 a^2 \sqrt{V_o}} \f{n^2-1}{n^2-4}
                \int d\Omega~\dgamma_{ij} \Pij \, , \\
    \pi^{(1)}_{nlm}  &=& 6 \sqrt{V_o} a^2  \int d\Omega~ \delta \tilde{\pi}^{ij} \Qij \, , \\
    \pi^{(2)}_{nlm}  &=& 6 \sqrt{V_o} a^2 \int d\Omega~ \delta \tilde{\pi}^{ij} \Pij \, .
\ea
Using the fundamental equal-time Poisson brackets $\{ \delta \gamma_{ij} (x), \delta \tilde{\pi}^{kl}(x')\}= \delta(x - x') \delta^k_{(i} \delta^l_{j)}$, one can derive the Poisson brackets for the amplitudes in the hyperspherical harmonics expansion:
\be
\left\{ \gamma^{(a)}_{nlm},  \pi^{(b)}_{n'l'm'} \right\} = \delta^{ab} \delta_{nn'} \delta_{ll'} \delta_{mm'} \, .
\ee
Similarly, it follows from $\{ \delta \phi (x), \delta \tilde{\pi}_\phi (x')\}= \delta(x - x')$ that
\be
\left\{ \delta \phi^{nlm},  \pi_\phi^{n'l'm'} \right\} = \delta_{nn'} \delta_{ll'} \delta_{mm'} \, .
\ee

\end{document}